# Euclid Near Infrared Spectrometer and Photometer instrument flight model presentation, performance and ground calibration results summary

Maciaszek T.: Ctr. National d'Études Spatiales, and LAM (Laboratoire d'Astrophysique de Marseille) UMR 7326 (France)
Ealet A.: Institut de Physique des 2 infinis de Lyon (Univ Lyon, Univ Claude Bernard Lyon 1, CNRS/IN2P3, IP2I Lyon, UMR 5822 (France)
Gillard W.: Aix Marseille Univ, CNRS/IN2P3, CPPM, Marseille, France
Jahnke K.: Max-Planck-Institut für Astronomie (Germany)
Barbier R.: Institut de Physique des 2 infinis de Lyon (Univ Lyon, Univ Claude Bernard Lyon 1, CNRS/IN2P3, IP2I Lyon, UMR 5822, (France)
Prieto E.: Aix Marseille Université, CNRS, LAM (Laboratoire d'Astrophysique de Marseille) UMR 7326 (France)
Bon W., Bonnefoi A., Caillat A., Carle M., Costille A., Ducret F., Fabron C., Foulon B., Gimenez JL., Grassi E., Jaquet M., Le Mignant D., Martin L., Pamplona T., Sanchez P.: Aix Marseille Université, CNRS, LAM (Laboratoire d'Astrophysique de Marseille) UMR 7326, Marseille, (France)
Clémens JC., Caillat L., Niclas M., Secroun A.: Aix Marseille Univ, CNRS/IN2P3, CPPM, Marseille, France
Kubik B., Ferriol S.: Institut de Physique des 2 infinis de Lyon (Univ Lyon, Univ Claude Bernard Lyon 1, CNRS/IN2P3, IP2I Lyon, UMR 5822 (France)
Berthe M., Barrière JC, Fontigne J.: Commissariat à l'Énergie Atomique (France)
Valenziano L., Auricchio N., Battaglia P., De Rosa A., Farinelli R, Franceschi E., Medinaceli E., Morgante G., Sortino F., Trifoglio M.: INAF Bologna (Italy)
Corcione L. , Capobianco V., Ligori S.: INAF - Observatorio Astronomico di Torino (Italy)
Dusini S., Borsato E., Dal Corso F., Laudisio F., Sirignano C., Stanco L., Ventura S.: INFN Padova (Italy)
Patrizii L., Chiarusi T., Fornari F., Giacomini F., Margiotta A., Mauri N., Pasqualini L., Sirri G., Spurio M., Tenti M., Travaglini R.: INFN Bologna (Italy)
Bonoli C., Bortoletto F., Balestra A., D'Alessandro M.: INAF - Observatorio Astronomico di Padova (Italy)
Grupp F., Penka D., Steinwagner J.: Max-Planck-Institut für extraterrestrische Physik (Germany)
Hormuth F., Schirmer M., Seidel G.: Max-Planck-Institut für Astronomie (Germany)
Padilla C.: Institut de Física d'Altes Energies (IFAE) (Spain)
Casas R., Lloro I.: Institut de Ciències de l'Espai, IEEC-CSIC (Spain)
Toledo-Moreo R., Gomez J., Colodro-Conde C., Lizán D.: Space Science and Engineering Lab (SSEL), Universidad Politécnica de Cartagena (Spain)
Diaz JJ; Instituto de Astrofisica de Canarias (Spain)
Lilje PB.: University of Oslo (Norway)
Andersen MI., Andersen J., N. Sørensen A.: Dark Cosmology Centre, Niels Bohr Institute, Copenhagen University (Denmark)
Hornstrup A., Jessen NC.: DTU Space, Denmark
Thizy C.: Université de Liège - ULg CSL (Centre Spatial de Liège)
Holmes W., Pniel M., Jhabvala M., Pravdo S., Seiffert M., Waczynski A.: NASA (USA)
Laureijs RJ, Racca G., Salvignol JC, Boenke T., Strada P.: European Space Agency/ESTEC
Mellier Y.: Institut d'Astrophysique de Paris (France) and Commissariat à l'Énergie Atomique (France)

on behalf of the Euclid Consortium

**ABSTRACT**

The NISP (Near Infrared Spectrometer and Photometer) is one of the two Euclid instruments (see ref [1]). It operates in the near-IR spectral region (950 –2020nm) as a photometer and spectrometer. The instrument is composed of:
– a cold (135 K) optomechanical subsystem consisting of a Silicon carbide structure, an optical assembly, a filter wheel mechanism, a grism wheel mechanism, a calibration unit, and a thermal control system,
– a detection system based on a mosaic of 16 H2RG with their front-end readout electronic, and
– a warm electronic system (290 K) composed of a data processing / detector control unit and of an instrument control unit that interfaces with the spacecraft via a 1553 bus for command and control and via Spacewire links for science data.
This paper presents:
– the final architecture of the flight model instrument and subsystems, and
– the performance and the ground calibration measurement done at NISP level and at Euclid Payload Module level at operational cold temperature.

## 1. INTRODUCTION

*Euclid* is a wide-field space mission concept dedicated to the high-precision study of dark energy and dark matter (see ref [15]). *Euclid* will carry out an imaging and spectroscopic wide survey of the entire extra-galactic sky (15000 $deg^2$) along with a deep survey covering at least 40 $deg^2$. To achieve these science objectives, the *Euclid* reference design consists of a wide field telescope to be placed in L2 orbit with a 6 years' mission lifetime. The payload consists of a 1.2 m diameter 3-mirror Korsch telescope with two channels: the visible instrument (VIS) and the Near Infrared Spectrometer and Photometer (NISP). Both instruments observe simultaneously the same Field of View (FoV) on the sky and the system design is optimized for a sky survey in a step-and-stare tiling mode.
The NISP Instrument is operating in the 950–2020 nm range at a temperature lower than 140 K, except for detectors, which are cooled down to ~95 K or below. The warm electronics is located in the service module at about 290 K.

The NISP instrument has two main observing modes: the photometric mode, for the acquisition of images with broad band filters, and the spectroscopic mode, for the acquisition of slitless dispersed images on the detectors.
In the photometer mode the NISP instrument images the telescope light in the wavelength range from 950 nm to 2020 nm ($Y_E$, $J_E$, $H_E$ bands). The spatial sampling is 0.3 arcsec per pixel. The FoV of the instrument is 0.55 $deg^2$ having a rectangular shape of 0.763 deg × 0.722 deg.
In the spectrometer mode the light of the observed target is dispersed by means of grisms covering the wavelength range of 930–1890 nm. In order to provide a flat resolution over the specified wavelength range, four grisms are mounted in a wheel. These four grisms yield three dispersion directions tilted against each other by 90° in order to reduce confusion from overlapping (due to slitless observing mode), but only 3 grims are used; 2 red grism with 0° and 180° dispersion and one blue grism with 0° dispersion. The field and waveband definitions used in the individual configurations for spectroscopy and photometry are:

- Three photometric bands (see ref [2]):
  1. $Y_E$ Band: 950−1212 nm
  2. $J_E$ Band: 1168−1567 nm
  3. $H_E$ Band: 1522–2021 nm
- Two Slitless spectroscopic bands:
  1. Red 0° and 180° dispersion: 1254−1850 nm
  2. Blue 0° dispersion: 920−1300 nm

The spectral resolution shall be higher than 250 for a homogenously illuminated object of one arcsec size. For such an object, the flux limit in spectroscopy shall be lower than $2x10^{-16}$ $erg{\cdot}cm^{-2}{\cdot}s^{-1}$ at 1600 nm wavelength. As with all slitless spectrographs, the real resolution varies with the object size (the smaller the size, the higher the resolution).
The image quality of the instrument in flight shall deliver a 50% radius encircled energy better than 0.3 arcsec and a 80% radius radius encircled energy better than 0.7 arcsec. There is a variation due to diffraction with wavelength.

The NISP budgets are presently the following:
The instrument sits in a box of 1.0 × 0.6 × 0.5 m
The total mass of the instrument is 155 kg
The maximum power consumption is 180 W
The instrument will produce 290 GBit of data per day

European Contributor countries for NISP are France, Italy, Germany, Spain, Denmark and Norway. USA (NASA) has provided the H2RG detectors, the cold proximity electronics and the detector-to-cold-electronic cold flex cables.

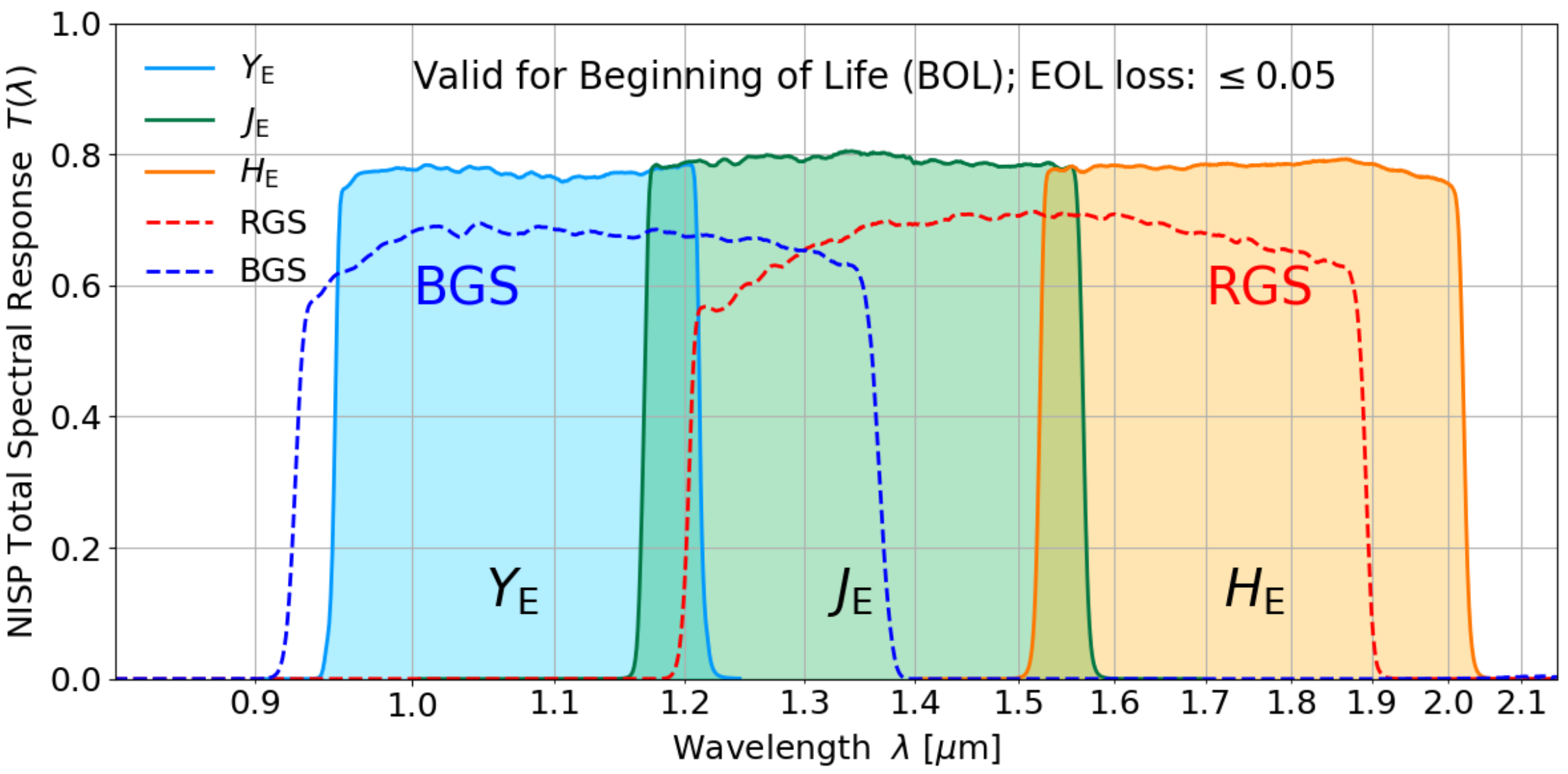

Figure 1: Total spectral response of the NISP photometric and spectroscopic channels, including all optical surfaces and detector quantum efficiency. Figure credit: Euclid Consortium.

# 2. NISP GLOBAL DESCRIPTION

The NISP instrument consists of three main Assemblies

- The Opto-Mechanical Assembly (NI-OMA), composed of the Mechanical Support Structure (NI-SA) and its Thermal Control (NI-TC), the Optical Elements Assembly (NI-OA), the Filter Wheel Assembly (NI-FWA), the Grism Wheel Assembly (NI-GWA), the Calibration Unit (NI-CU). The NI-OMA structure again supports the NI-OA, NI-CU, NI-FWA and GWA, and the detection system (see below). It provides the thermo-mechanical interface towards the Euclid Payload Module (PLM).
- The Detector System Assembly (NI-DS) is composed of the Focal Plane Assembly (NI-FPA) and the Sensor Chip System (NI-SCS). The NI-FPA provides the structural elements of the of NI-DS, that accommodate and provide mechanical support to the Sensor Chip System (NI-SCS). The NI-SCS is composed, in turn, of 16 H2RG infrared detectors, the associated 16 ASICS (Sidecars), passively cooled at operating temperature via the radiators (<100K for the detectors; 140K for the ASICS Sidecar) and 16 low thermal conductance, low emissivity copper/constantan flex cables that are used to connect each H2RG/ASIC pair. Thermal stabilization of the detector is "naturally" obtained thanks to the very good thermal stability provided by the Euclid PLM at the NISP interfaces.
- The Warm Electronics Assembly (NI-WE), composed of the Data Processing Unit and Control Unit (NI-DPU/DCU), and the Instrument Control Unit (NI-ICU). The NI-ICU is managing the commanding and the control of the instrument. It is interfaced with the satellite via a 1553 bus. The NI-DPU/DCU controls the NI-SCS and basic image processing such as co-adding (DCU function) and the science onboard data processing, the compression and transfer of scientific data to the S/C Mass Memory using Spacewire links (DPU function). The NI-DPU/DCU functions are regrouped in a single mechanical box for controlling eight detectors. There are two NI-DPU/DCU boxes.

The NI-DS is screwed on the NI-OMA (SiC panel to SiC panel). The NI-OMA+NI-DS is located in the Euclid spacecraft Payload module in a cold environment (130K). The NI-WE is located in the Euclid spacecraft Service Module at “room temperature” (290 K). A long-dedicated harness (6m) interconnects the NI-OMA, the NI-DS, the NI-WE, and different spacecraft electronics boxes.

The instrument has a specific observing sequence that will be repeated throughout the mission lifetime.
The survey is decomposed into fields (see ref [16]). To avoid confusion and increase spatial resolution, each field is observed with 4-dithered frames for each band. Only one grism will be used at each dither. The spectrum confusion is minimized thanks to the use of two grisms being used at –4°, 0°, 180° and 184° dispersion direction. During the nominal wide survey, only two red grisms are used. Meanwhile, the blue grism is used only during the deep survey.
Each sequence of exposures consists of a 565 s spectroscopic observation spectroscopic observation followed by 112s of photometric observations in each of the $Y_E$, $J_E$, $H_E$ photometric bands. The Filter and the Grism Wheels are activated between each observation to set the instrument in the proper configuration. Due to the specificity of the H2RG detector, no shutter is needed to stop the integration. An opaque plate is included in the filter wheel for dark calibration purposes, and to prevent persistence effects due to charge accumulation during satellite slews.

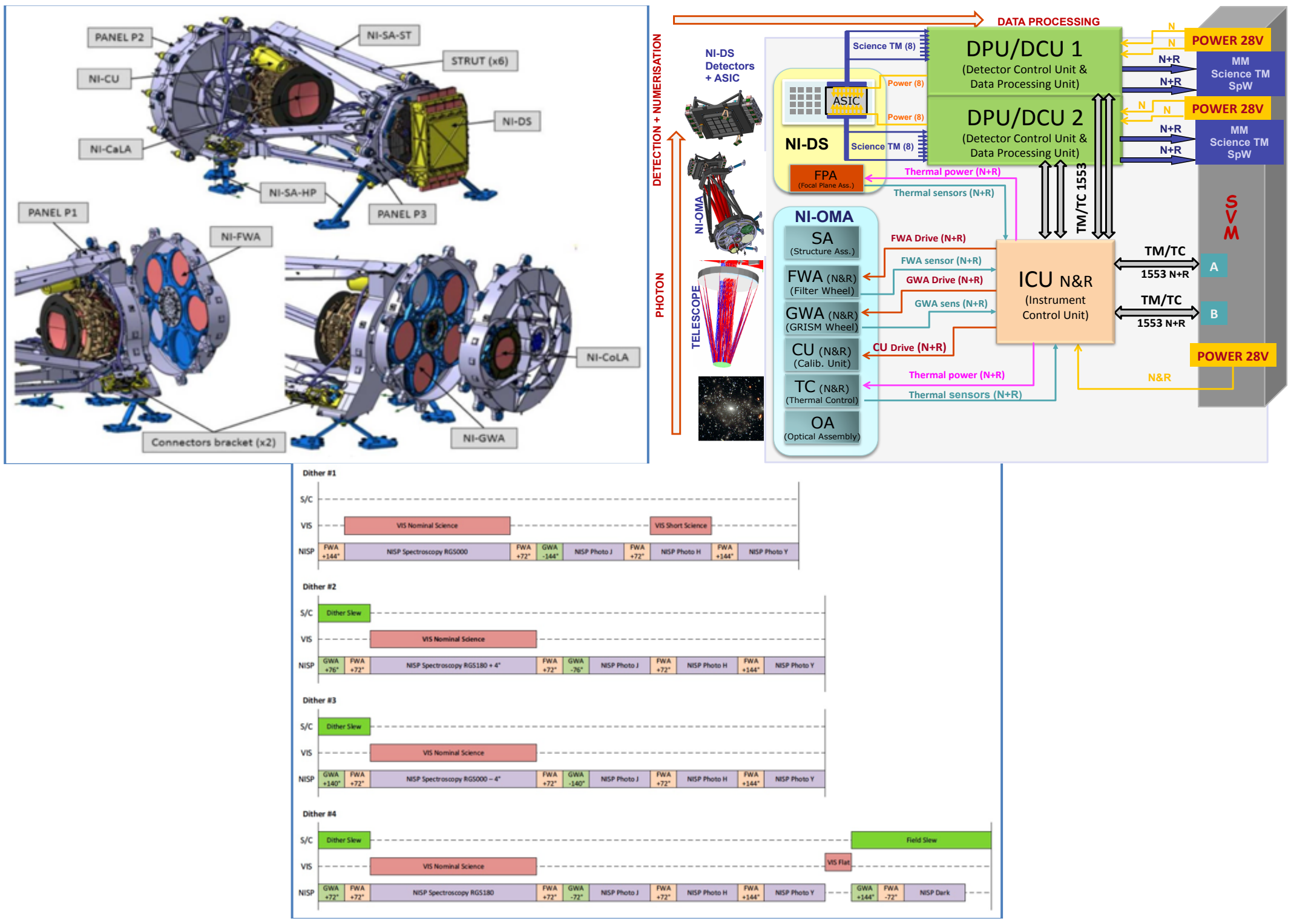

Figure 2:: NISP global overview, functional scheme, and observing sequence (wide survey)

## 3. NISP OPTO-MECHANICAL ASSEMBLY (NI-OMA)

**The structure (see ref [3]):**

The structure is made of Silicon Carbide (SiC). This choice has been the result of a long trade-off among carbon fiber, aluminum and SiC, mainly. The main driver was the very tight constraint of dimension stability of the system from AIV to the end of the mission. The SiC structure is interfaced with the Euclid PLM Baseplate via an Invar Hexapod.

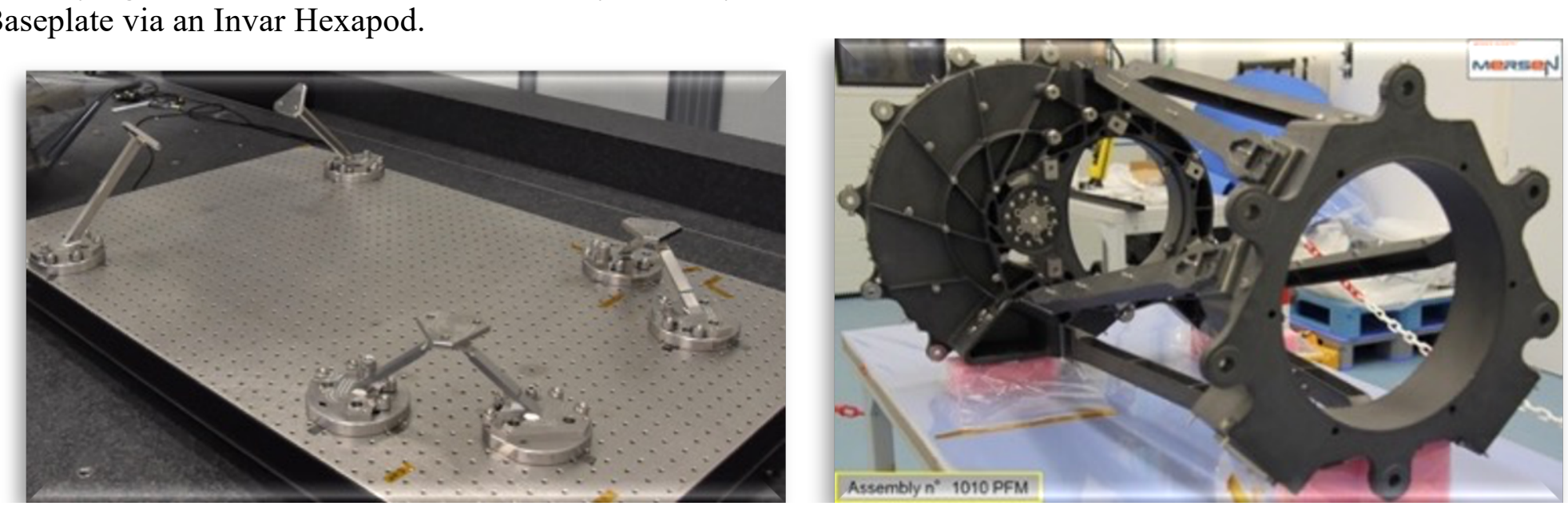

Figure 3: NISP Hexapod (FM) and SiC Structure (FM)

**The thermal control:**

Operating temperature and thermal stability of the units are key drivers for the instrument. The structure (NI-SA) is by design thermally insulated from the PLM baseplate through Invar bipods and monopods (i.e. NISP Hexpod). The total conductance from NI-OMA to the Baseplate is approximately 0.035 W/K, which minimizes the transferred heat between these elements. In this configuration, the units inside the NI-OMA, and especially the optical lenses, are less sensitive to fluctuations and can exploit the whole instrument thermal mass to operate in a more stable condition. Cooling down to the required operating temperatures is performed, mainly, by heat extraction through two conductive thermal interfaces, connected to radiators, provided by the PLM located on the NI-DS. The structure is entirely made of SiC which, given the good thermal conductivity of such material, ensures good temperature uniformity and an efficient heat extraction.

This NI-OMA is operated at a temperature around 135 K with a stability better than 0.3 K for *all* the life duration including ground tests when performance are measured for the full mission operation, from ground to end of life.

Radiative loads from the PLM cavity environment are efficiently shielded by an MLI shroud that surrounds the whole instrument. Both blanket surfaces, internal and external, are made of black Kapton to minimize straylight contamination.

The figure below shows the results of the NISP FEM thermal model with the external interfaces in warm conditions and units nominal dissipation. The NI-SA temperature is around 132 K with an overall thermal gradient below 2 K and the optical lenses showing temperature differences of less than 0.5 K. The detectors front end electronics is the warmest unit (~139 K), due to its active dissipation, while the FPA operating temperature is only 0.5 K higher than its reference interface set at 95 K.

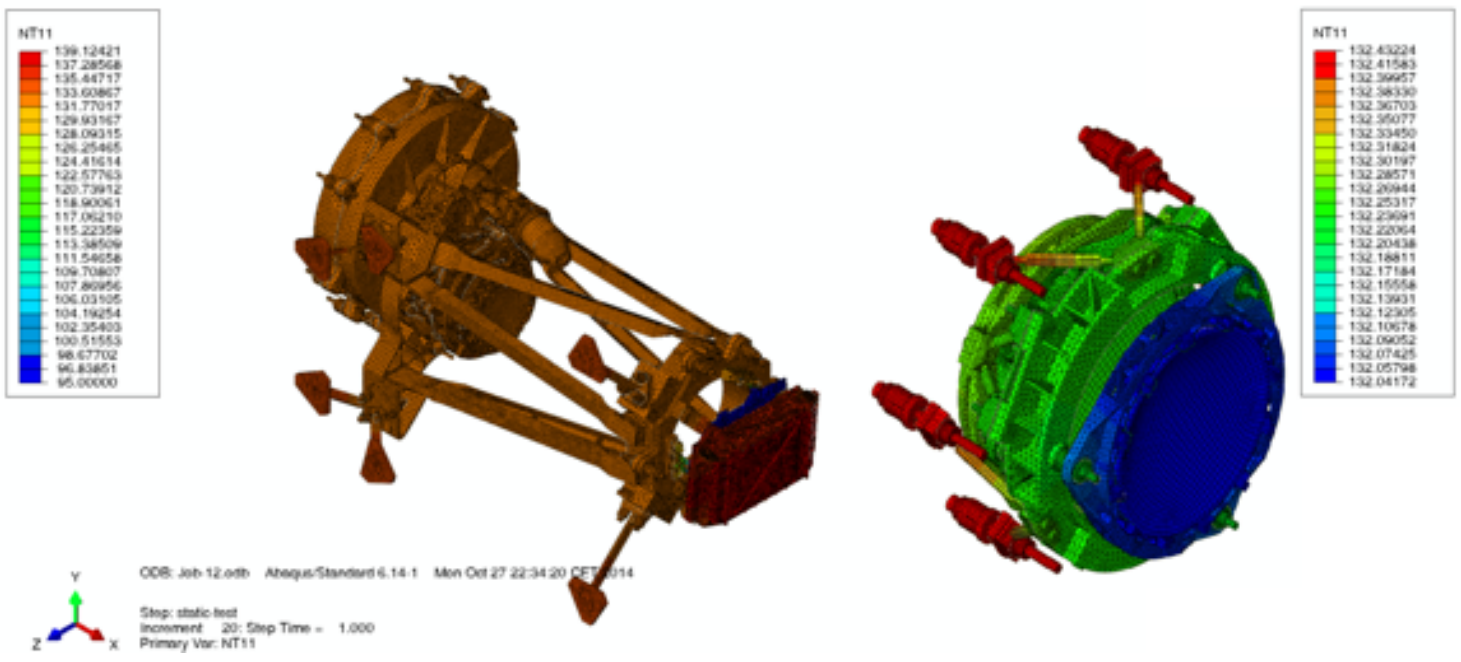

Figure 4. NISP FEM thermal model results: the NI-OMA (left panel) and the CaLA unit with the lenses (right panel)

**The Optics (see ref [4]):**

Main function of the NI-OA is the accommodation of the 4 lenses L1, L2, L3 and corrector lens L4. The first 3 lenses (L1–L3) are summarized in the Camera Lens Assembly (CaLA), while the single corrector lens is allocated in the Collimator Lens Assembly (CoLA).

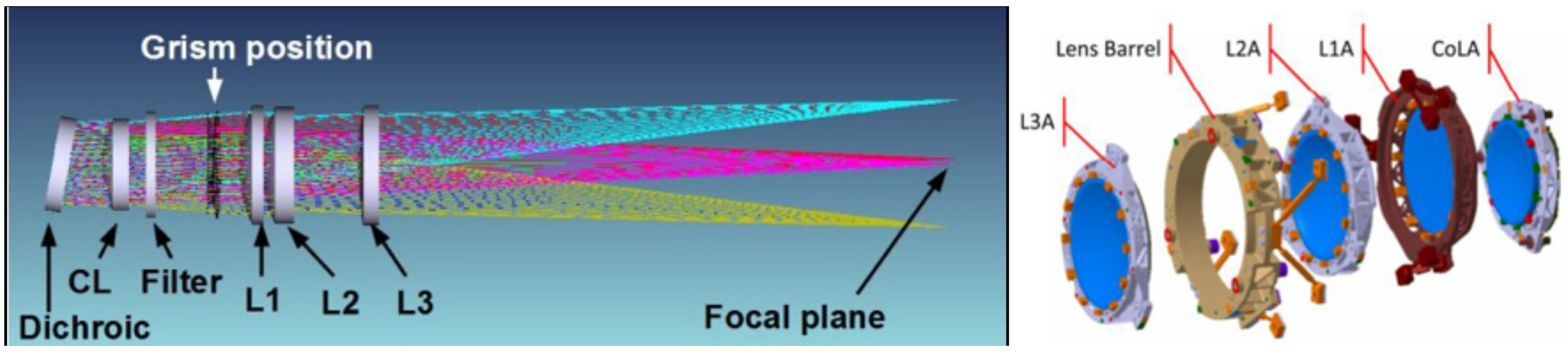

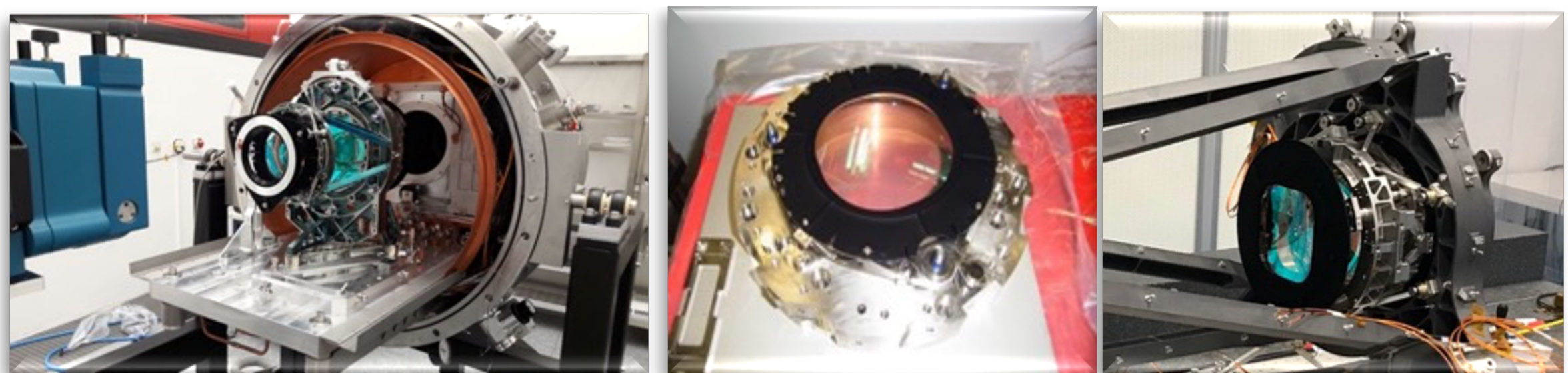

Figure 5: NISP Optics (Flight Model, FM)

The optical system operates in a temperature range between 132 and 134 K. Consequently, the lenses need to have different shapes during manufacturing, to reach the desired shape at operational temperature. For the “cold – warm” calculation, the warm lens geometry has to be derived from the ideal “cold” optical design, which is valid for one specific temperature of the operational temperature range. The transfer of the lens parameters from OPS (operational conditions) to RT (“warm” room temperature) is dependent on the temperature dependency of material and optical parameters.

At room temperature, the lens material constants are well known, however at the operational temperature range the most important parameters for the optical design, such as refractive index, CTE were only partially available for the used lens materials. The mechanical lens material properties such as CTE, strength, etc. have been measured down to the operation temperature range.

The design drivers for the adaption rings (AR) are high precision, cryogenic operation temperature (110 K) and the large dimension of the lenses (168 mm). The design concept of the ARs is based on flexure hinges, which provide sufficient protection against vibration loads at ambient temperature as well as high precision (< ±10 µm) and stability at cryogenic temperatures.

Criteria for the flexure hinge design are the low radial forces at cryogenic conditions to avoid any refractive index and polarization variations. The design is compliant with the large temperature differences between assembly and operation, the high precision and low-deformation requirements of the lenses, as well as with the deviating CTEs of the selected lens materials.

Each lens is glued (epoxy bond) in an adaption ring via a double pad, which provides the necessary elasticity caused by different CTEs of the lens and ring materials, as well as it allows a high position accuracy of the lenses relative to the lens barrel and the optical axis. The double pad itself is bonded to the lens using the same glue. The glue pad dimensions are the same for each spring with an accuracy <0.1 mm respectively. Otherwise, asymmetric deformation of the corresponding lens is introduced after cooling down, and hence, the accurate position and form of the lens cannot be guaranteed. Also the high precision manufacturing process of the springs by means of wire eroding is of critical importance, since its thickness shows up with the third power in the spring force, which introduces additional lens movement at operational temperature of ~135 K.

Baseline for AR material selection is the similarity of the CTE of both lens and AR to minimize any stress in the lens material. The remaining force at OPS temperature is further reduced by the springs and so the lens deformation is kept as low as possible. Benefit of the design is that the assembled AR is mechanically very rigid and withstands the vibration loads at RT.

The assembled ARs of the CaLA lenses are mounted into the lens barrel, which provides high precision position of the lenses. The LB I/F the unit is decoupled from the SiC structure of by using 3 bipods.

The performance of the flight model are extremely good, better than the already very stringent requirements; (see ref [4])

**<u>The Filter wheel:</u>**

The Filter Wheel Assembly (NI-FWA) is made of: A Cryo-mechanism, three filters, an “open” position for which the optical beam can cross the FWA without interference, an opaque “close” position for which the optical beam is blocked, and a wheel structure in Invar. The operational temperature is around 135 K.
The performance of the flight model is fully compliant with the requirements.

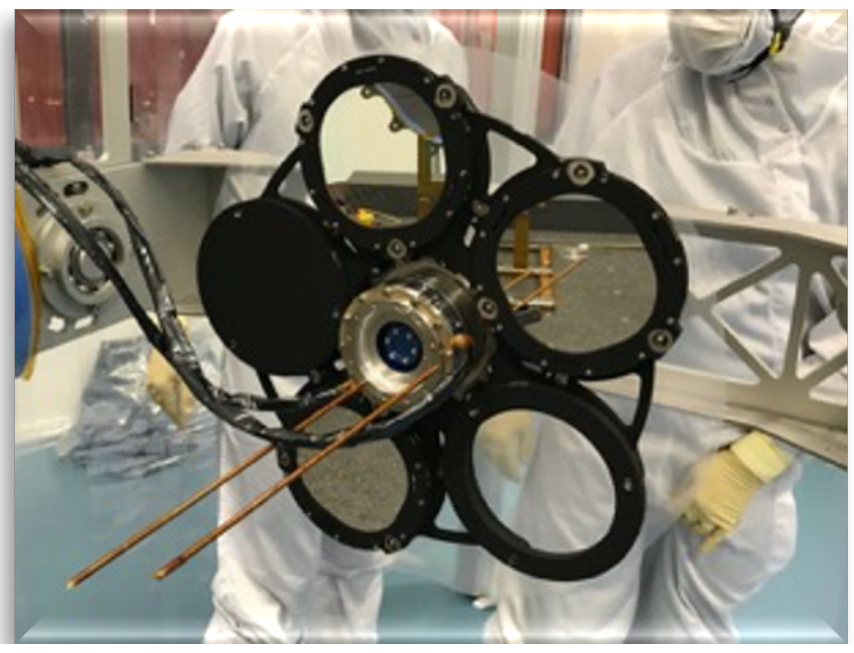

Figure 6: NISP Filter wheel (FM)

**The Grism wheel:**

The Grism Wheel Assembly (NI-GWA) is made of: A Cryo-mechanism, four Grisms, an “open” position for which the optical beam can cross the GWA without interference, and a wheel structure in Invar. The operational temperature is also around 135K.
Again, the performance of the flight model are all compliant with the requirements.

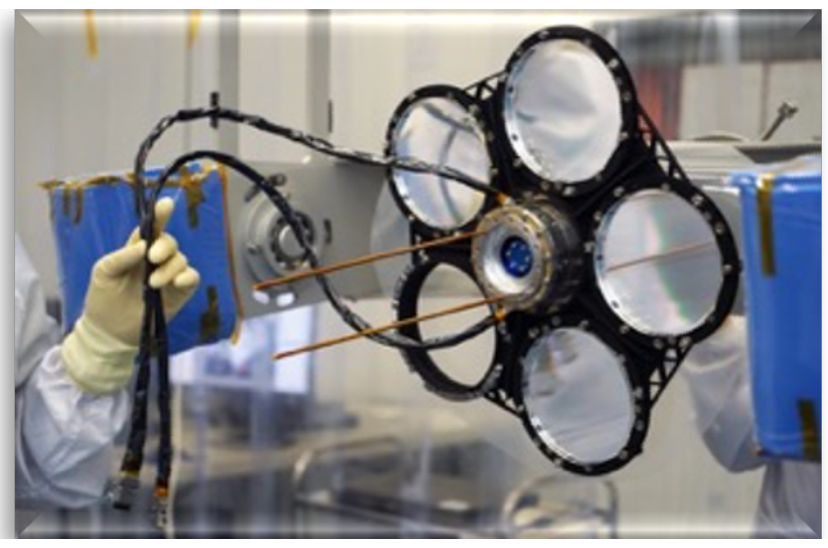

Figure 7: NISP Grism wheel (FM)

**The Cryomechanism (see ref [5]):**

The FWA and the GWA are motorized with two identical Cryo-Mechanisms (CM). The CM operates from room temperature down to cryogenic temperatures (130 K). It includes a stepper motor that performs a coarse positioning, rotating the wheel at any of the 360 positions within +/–0.3° of uncertainty (due to the bearings’ friction). When arrived at the required position, the motor detent torque (40 mN·m) combined with the friction torques are enough to maintain the wheel position. During this motion, five degrees of freedom (DOF) are locked by the bearings assembly. The only DOF which is free is the rotation around the bearing axis. The cryomechanism is powered only while actuation is required, when not operated, the cryomechanism is fully OFF.
The performance of the flight model are all compliant with the requirements and particularly, the life time duration has been fully qualified.

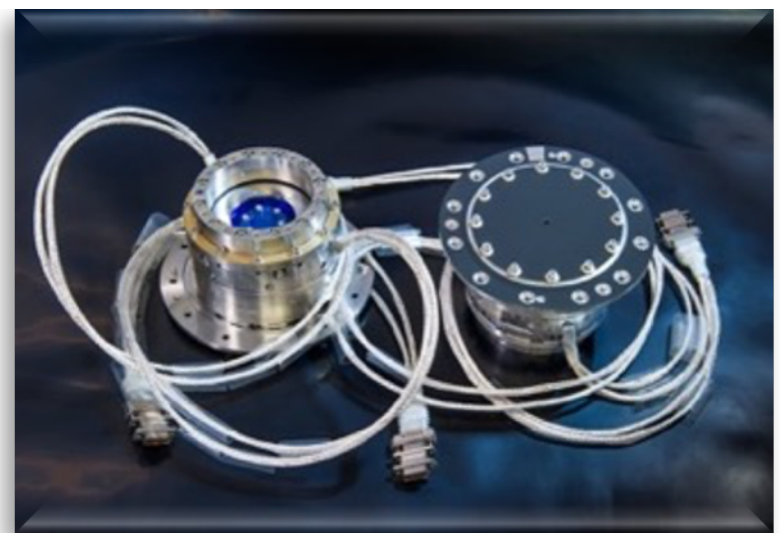

Figure 8: Cryomechanism FM

**The Grisms (see ref [6]):**

Four grism are mounted on the GWA, three “red” with a spectral bandpass [1250–1850 nm] and one “blue” with a spectral bandpass [920-1300 nm]. The four grisms have about 14 grooves/mm.
Each NI-GS is composed of the grism itself (the optical element) glued on an Invar mechanical mount. A baffle is mounted on the mount.
The grism itself combines four optical functions in one component:

- A grism in Suprasil 3001 made of a grating engraved on the prism hypotenuse to make the light un-deviated at a chosen wavelength. In addition, a spectral wavefront correction is done by the curvature of the grating grooves
- A spectral filter done by a multilayer filter deposited on the first surface of the prism
- A focus function done by the curvature of the first surface of the prism

The optical part of the grisms is glued in a mechanical Invar M93 ring through 9 flexible blades that compensate the small CTE difference between Suprasil 3001 and Invar. Three blades enable to minimize stresses in the optical element due to thermal differences (from 300 K to 130 K) and to interface defaults.

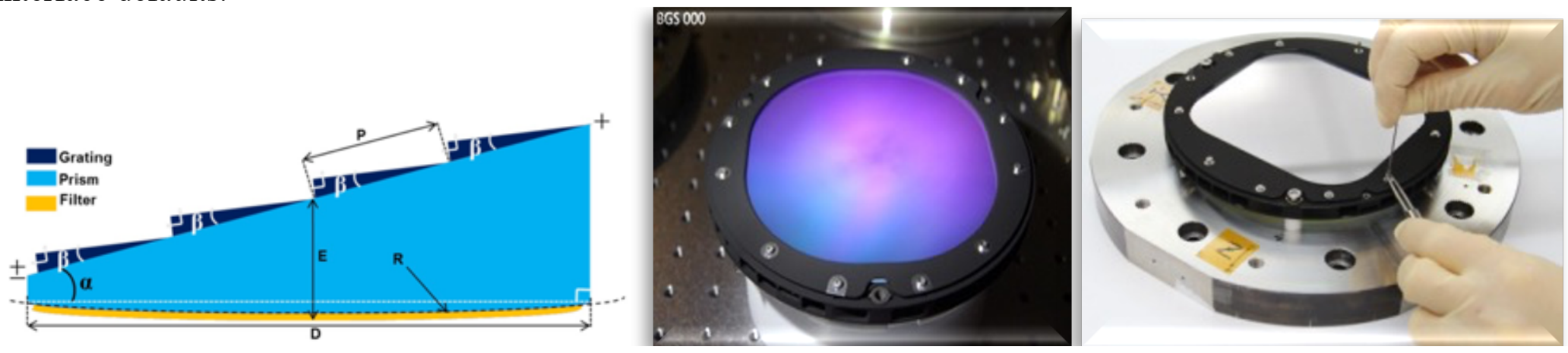

Figure 9: Grism (FM)

**The Filters:**

The three infrared filters ($Y_E$-, $J_E$- and $H_E$-band) for the NISP instrument are realized as double-sided interference filters coated with the PARMS (plasma assisted reactive magnetron sputtering) process. Each side of the ~12 mm thick Suprasil 3001 filter substrate is coated with a stack of up to 200 individual layers, resulting in a stack thickness up to 20 µm per side.

With a clear aperture of 126 mm (total diameter 130 mm) this requires the use of coating machines designed for 8-inch substrates in order to achieve the required coating thickness homogeneity and to reduce the resulting transmissive wavefront error.
As the filters are not simply flats but have a lenticular shape, with ~10 m focal length, manufacturing and verifying the uncoated substrates alone is complex enough, but furthermore the circumference (i.e. sides of the filter substrate) has to be polished down to ~2 nm RMS roughness while retaining excellent circularity. This is necessary to allow controlled gluing of the filters into their mounts, later to be integrated into the FWA.
In previous studies we have also investigated the possibility to use the more conventional ion-assisted deposition (IAD) coating approach. While this process results in slightly better cosmetics of the filter surfaces, PARMS allows for better thickness homogeneity and was chosen for that reason. In the run-up to final filter production we have produced IAD prototypes of the $H_E$-band filter as well as a down-scaled PARMS prototype of the $Y_E$-band filter.
Both test productions have resulted in well reproduced transmission properties, i.e. the transfer of the theoretical design with extended blocking and transmission >95% in the passband worked very well.
Thermal shift of the bandpass between room temperature and operational temperature (~130 K) has been reproducibly measured to be <2 nm and well predictable. Measurements of the substrate bending due to internal coating stresses have been found to be in agreement with model predictions and within tolerance budget.

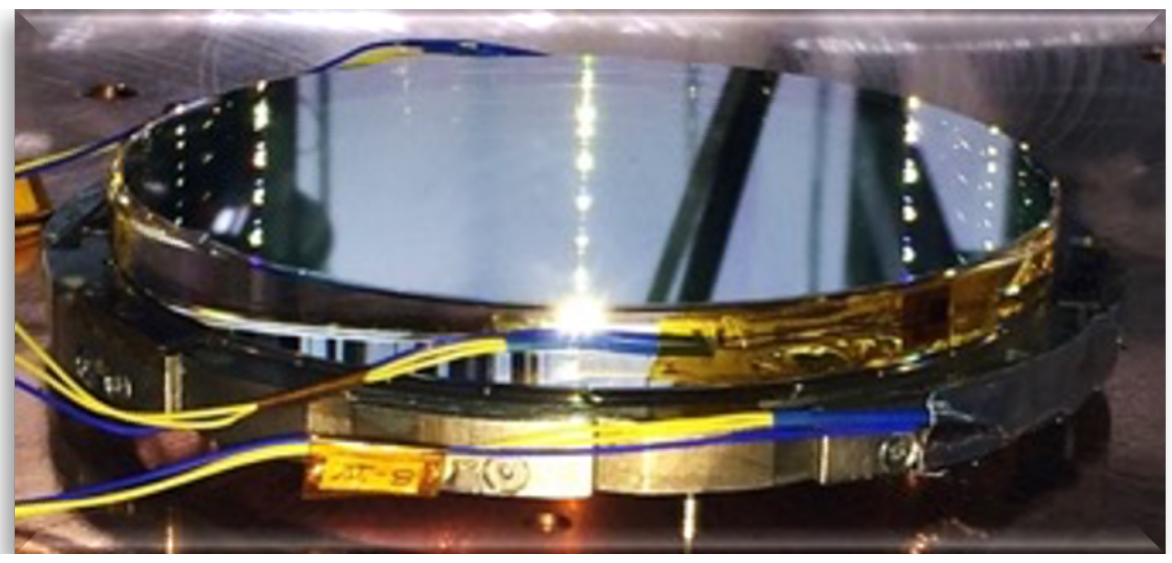

Figure 10: A Filter FM

**Calibration Unit:**
The NISP Calibration Unit allows in-flight calibration of the infrared detector array. The unit provides stable illumination of the image plane at five different infrared wavelengths, allowing for small-scale flat field calibration and measurements of the detector linearity.
The design is relatively simple with 2×5 LEDs (one nominal and redundant per wavelength) inside the calibration unit pointing to a small patch of Spectralon PTFE material. The Lambertian scattered light is directly pointed towards the detector through a set of baffles without going through any of the optics.
Control of the LED brightness and thus received flux is performed by current and duty cycle regulation of the drive signal in the instrument control unit (ICU).

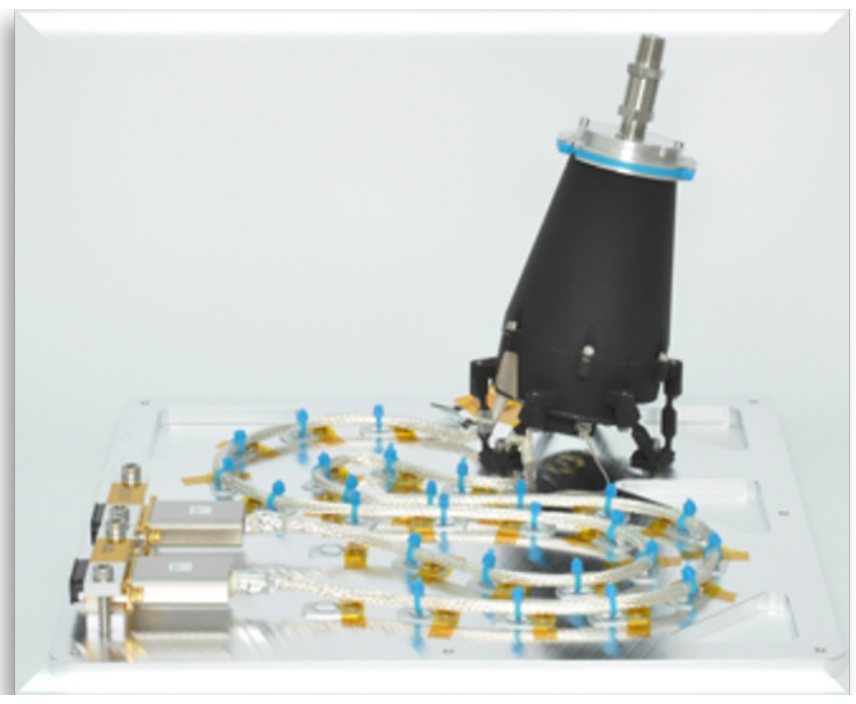

Figure 11: Calibration source FM.

## 4. THE NISP DETECTOR SYSTEM (NI-DS)

This assembly has the function to acquire the images by sampling the Field of View with an array of 4×4 NIR sensors hybridized on multiplexers (18µm pitch or 0.3 arcsec on the sky) and read out by the Sidecar ASICs. It is sequenced and read out by the NI-DPU processing to deliver digitalized data to the NI-DPU.
The NISP Detector System (NI-DS) is composed of:

1. A SiC panel called P4 (to be screwed directly on the SiC structure of the NI-OMA)
2. A Cold Plate (CSS) that supports the mosaic of 4×4 detectors. The Cold Plate is made of molybdenum and is held by three titanium bipods on the P4 Sic panel. A baffle (for detector protection), also made of molybdenum, is fixed on the CSS
3. A support structure for the Sidecars (SSS). It is fixed onto the panel P4 by three bipods made of Invar. The NI-SSS is made of aluminum
4. The Sensor Chip System (SCS), composed of the H2RG sensor with 2.3 µm cut-off (SCA), its cryo-flex cable (10 cm) and its ASIC sidecar electronic (SCE)

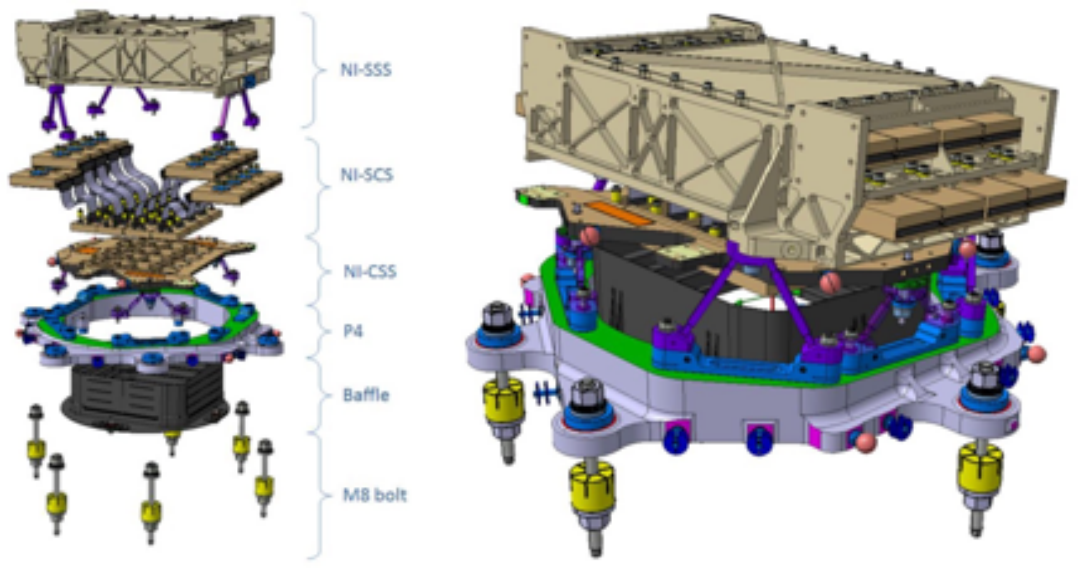

Figure 12: Focal plan overview

The operating temperature of the detectors (SCA) is lower than 100 K while each individual readout electronic (SCE) operates at around 140 K (ref [7]). Since the instrument units facing the detectors are controlled at a temperature below 135 K, the resulting thermal emission up to 2.3 µm ensures a very low thermal noise level. This configuration allows the optimization of the system thermal load on the satellite radiator and complies, at the same time, to all specifications in terms of noise.

SCA/SCE operation synchronism, at the level of a single master clock period (10 MHz, 50 nS allocated maximum differential skew budget), is important for a mosaic made of tightly coupled detectors with potential electrical crosstalk. It is ensured partly by SCE firmware specifically developed for the NISP application (see next paragraph) and partly by specific HW in the DCU electronics. Basically, all the SCE systems are driven by a common master clock and all writings to the SCE internal registers (configurations and command directives) are synchronized by shift-registers clocked by the same master clock and started by a common pulse.
Data and power supply connections to each SCE are done by an unusually long double-shielded cable harness. The length is primarily dictated by internal distribution inside the payload module and the service module where warm electronics boxes are accommodated and by thermal decoupling reasons.
The flight detector SCA characterization has been conducted (with SCE Engineering Models, EM) to obtain precise characteristics for noise, dark current, conversion gain, non-linearity of the pixel response, QE, inter-pixel capacitance crosstalk, full-well capacity, and persistence/latency (ref [8] and [9]). Complementary characterization tests with the complete SCS flight models (SCA, flex, and SCE flight models) have been performed at instrument level.
The performance of the flight model are all compliant with, and actually better, than the requirements.

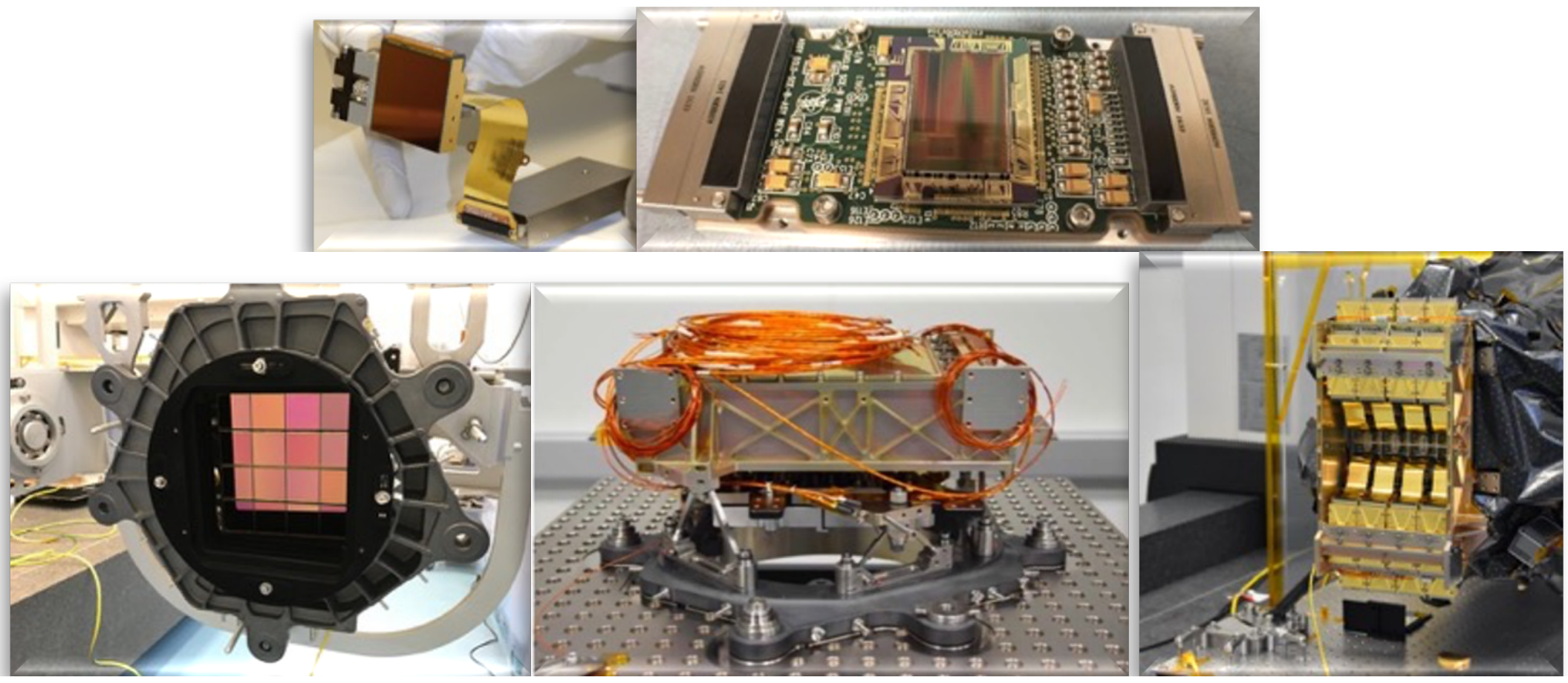

Figure 13: Views of the detector chip (SCA FM), electronic (FM) and NI-DS FM

All of these subsystems have been integrated and aligned on the structure to give the Optomechanical Assembly flight model.

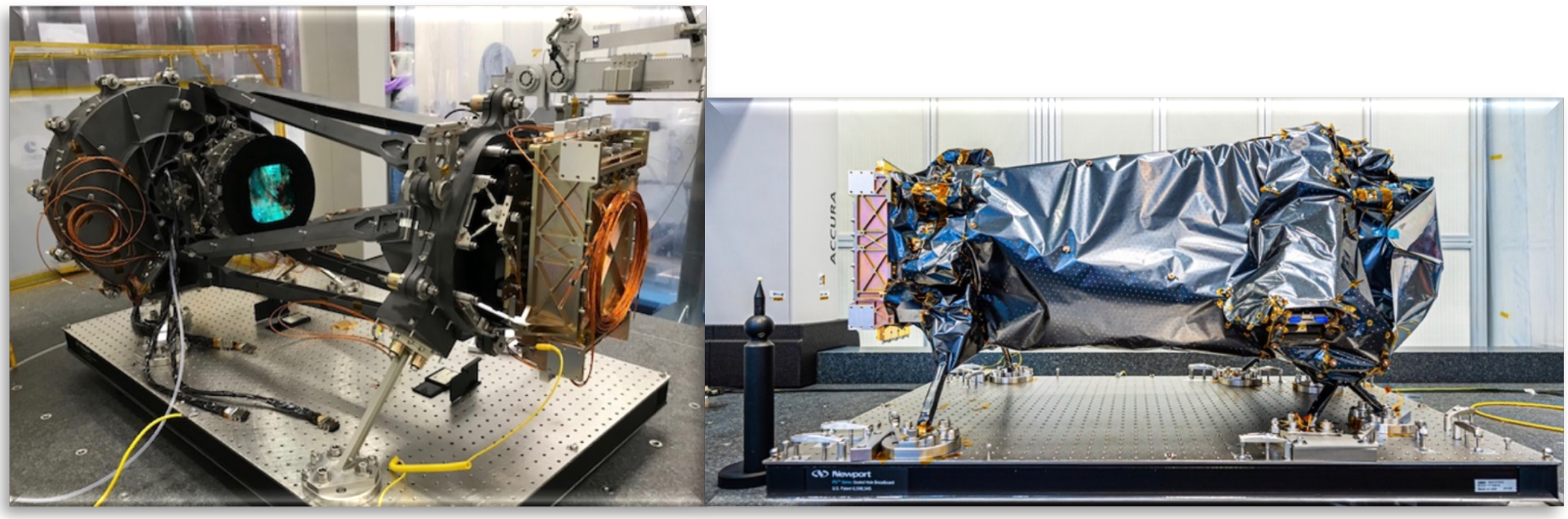

Figure 14: Views of the Optomechanical Assembly FM

## 5. THE NISP WARM ELECTRONICS

The NISP warm electronics is composed of two Data Processing Unit (NI-DPU) and one Instrument Control Unit (NI-ICU). The warm electronics is located in the service module of the spacecraft. A harness under Prime contractor responsibility (Airbus and Thales) provides the link with the NI-OMA and NI-DS. This cable will carry the LVDS signal for scientific data, housekeeping signals, control command and power supply for equipment.
The main challenge of the warm electronics is to process the amount of data delivered by the detector during the integration of the following frame.

**Data Processing Unit (NI-DPU):**
Each DPU unit is mounted around a shared Compact PCI bus structure with the exception of the main power supply system and the 8x DCU boards, each one managing one SCE/SCA detection pair. The two NI-DPUs are both including:

- 8x Detector Control Units (DCU) that provide clock and power to the readout electronics. In addition, these units will preprocess the data using FPGA boards
- Central Processor Unit that finalizes the on board data processing, compress and format the data sending them via SpaceWire link to the central spacecraft memory

Each DPU is hosting the following boards:

- CPCI Data Processors based on a Maxwell SCS750 board
- CPCI Data Routers
- CPCI Data Buffer
- Power Supply

Except the 8 DCUs, all boards in the DPU are in cold redundancy. Each DCU receives the data of one 2K×2K detector from one SIDECAR and performs the low level pre-processing foreseen in HW consisting of:

- Group of frames averaging
- Telemetry Extraction
- Extraction of sub-sets of programmed raw detector lines to be used on ground for monitoring purposes
- Co-added Frame data buffering and Spacewire transmission to Data Buffer Boards

At this interface level redundancy is supported by the full duplication of DPU hardware. Averaged data groups can be configured to be transmitted to one of the Data Buffer Boards available in each DPU, this is accomplished by duplication of the 8x Spacewire (SpW) links. The same redundant configuration is available at each DCU TMTC interface bus: the control link based on the RS485 standard can be configured to be driven by one of the two available CPCI data router boards.
The Data Buffer board allows the storage of up to 46 averaged frames per dither per detector with Telemetry and ancillary data from the 8x handled detection channels in double-buffering mode to ease the further data processing.

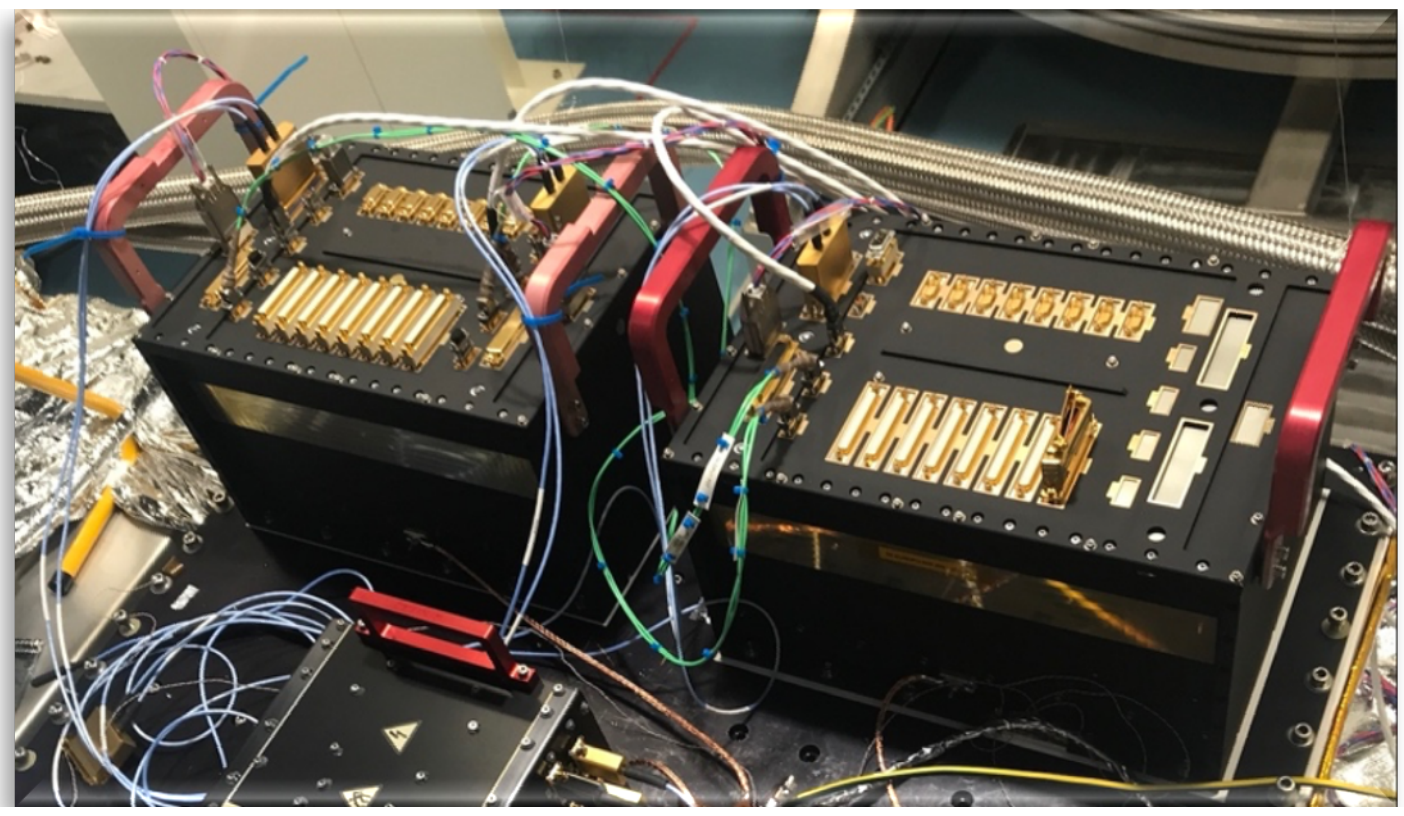

Figure 15: The two DPU FM boxes

**Instrument Control Unit Hardware (NI-ICU):**

- One Instrument Control Unit (NI-ICU) in charge of:
    - Interface with the spacecraft via a 1553 bus for the commanding of the NISP
    - Housekeeping management
    - Power supply (internal, Cryo Mechanism, Heaters, NI-CU)
    - Command signal to the cryo-mechanism, to the 5 LEDs calibration source and to the NI-OMA and NI-DS heaters (heater constant power is applied in open loop with power setpoint determined by ground operators)

The ICU has two sections (nominal and redundant), which are identical and operate in cold redundancy. Each NI-ICU (N or R) is divided in three boards, all of them interconnected by means of a backplane motherboard:

- LVPS (Low Voltage Power Supply): provides DC/DC converters to generate all the necessary secondary power supplies, as well as the 1553 transceivers for the NI-DPU link (1553 controller logic is actually located in the CPDU board, see next item).
- CDPU (Central Data Processing Unit): contains a LEON2-FT CPU embedded in a MDPA ASIC, which manages all the functions of the NI-ICU. This module also includes a RTAX FPGA that extends the functionalities of the MPDA, with the main aim of interfacing with the DAS module (next item). The 1553 transceivers for the S/C link and the test connector are also located in this board.
- DAS (Data Acquisition System): this board features all the analog acquisition and driving electronics that are used to interface with the rest of the NISP instrument, including the filter and grism wheels, heaters, temperature sensors, and calibration LEDs.

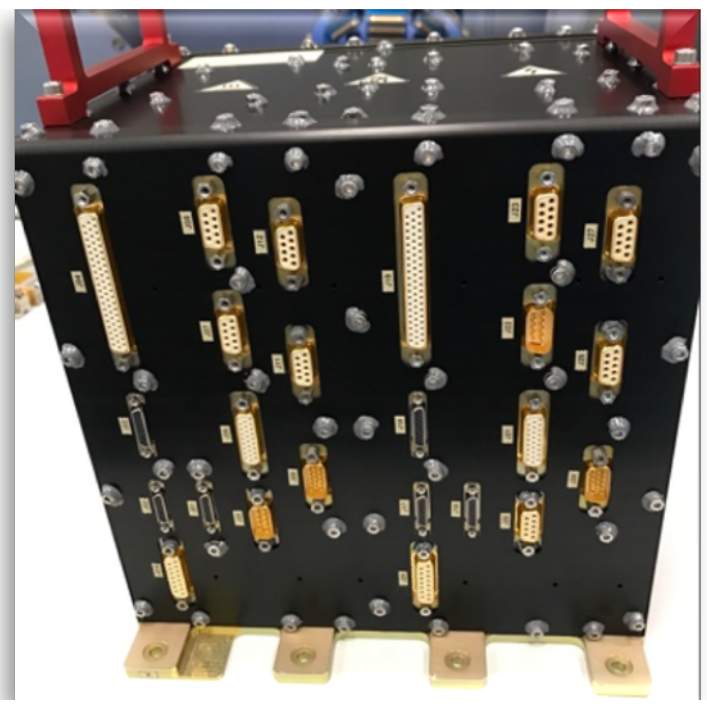

Figure 16: NI-ICU FM

**Instrument Control Unit Hardware Application Software see ref [10]):**

The ICU Application SW (ASW) is devoted to manage the satellite/platform interface, the ICU/DPU interface and all the functionalities related to instrument commanding. It is in charge of the following functions:

- TM/TC exchange with S/C CDMU on Nom/Red 1553 link
- TC decoding and distribution to NISP instrument subunits: NI-FWA electronics, NI-GWA electronics, NI-CU electronics, NI-TC electronics, NI-DPU/DCU/SCE
- Global instrument monitoring and HK packet generation
- Time management, propagation of the On Board Time (OBT) to the DPUs, TM time tagging and high level instrument internal synchronizations
- NISP operating mode management
- Execution of autonomous functions and FDIR algorithms and processes.
- Control of the calibration unit (ON/OFF, intensity level and current absorption handling)
- Control of filter wheels (reference position, position switch)

- Thermal control (open loop) of the NI-FPA detector cold-plate through temperature sensors and heaters
- High level handling of macro-commands submission to detector system
- Thermal control of the NI-OMA through temperature sensors and heaters
- Management of software maintenance, memory patch and dump (EEPROM patching is performed by the Boot SW)

The ICU ASW is based on RTEMS real-time operating system, in the space-qualified version by EDISOFT.
Telecommand and Telemetry packets will be based on the Packet Utilization Services (PUS) standard, with the implementation of services tailored to the specific needs of the Euclid project.
A coordinated effort is in place with the Prime of the Spacecraft and with the VIS CDPU ASW team in order to ensure a common approach and, as far as possible, implementation of services between NISP and VIS, so that the SW interfaces with the Spacecraft can be simplified and standardized.
The interface with the DPU is based on a second MIL-STD-1553 bus, similar to the one used between the Spacecraft and the Euclid instruments, in which the DPUs are configured as Remote Terminals and the ICU as the Bus Controller. The SW interface and communication protocol is an internally defined one, with the aim of reducing as much as possible the load of management tasks on the DPU processor, since this resource is needed for the demanding data processing tasks. The ICU ASW will decode the PUS formatted high level TCs and implement the low level sequences towards the two active DPUs.

## 6. ONBOARD DATA PROCESSING (SEE REF [11] AND [12])

The routine science NISP operations foresee 20 fields of observation per day, each one composed of four dithers where four exposures (one in spectroscopic mode, three in photometric mode, see above) each are taken, for a total maximum assigned science data telemetry of 290 Gbit/day. A dark exposure is taken during the spacecraft slew. This limited amount of allowed telemetry, together with the huge number of frames typically produced by IR detectors operated in multi-accumulation mode, have as a consequence the need to perform part of the processing pipeline directly on-board and to transfer to ground only the final products for each exposure. Moreover, final data must be also compressed to fit with the assigned telemetry throughput. A number of readout modes have been envisioned for the NISP instrument in the various development stages. Multi-accumulation (MACC) is the reference mode of operation for both spectrograph and photometer readout. MACC readout is a peculiar Up the Ramp process (UTR) where detector readouts are grouped in contiguous sets of readouts uniformly placed along the accumulated charge ramp. The data processing can be split into two main stages: stage 1 is implemented in the NI-DCU, directly interfaced to the SCS, where the first static basic pre-processing steps are performed, while stage 2, performed in NI-DPU, is devoted to the processing and compression of the final data frames.

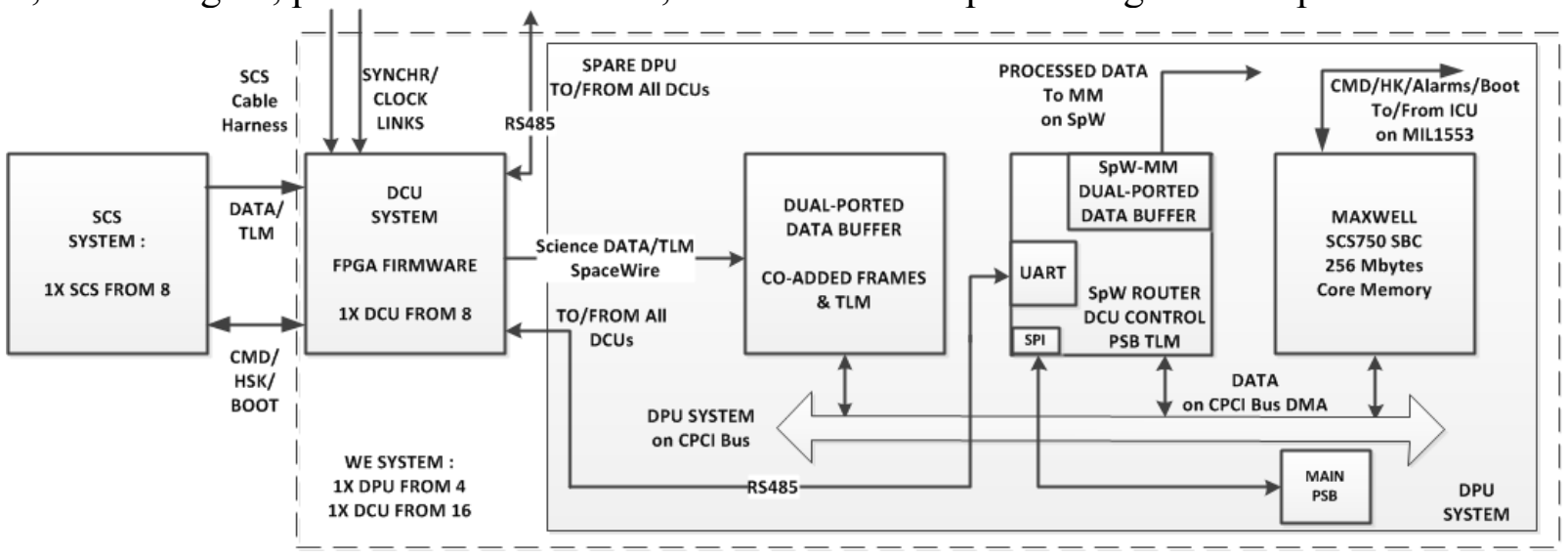

Figure 17: Pre-processing HW structure connected to 1× SCS single pair (H2RG SCA + SCE) from a total of 16× located inside the SCS system

The software architecture is dictated by the science requirements and depends on the hardware organization, in terms of DPU power, internal memory, available links with both DCU and SVM. During the previous different phases of the project various processing possibilities were analyzed, in terms of computational complexity, DPU internal memory needs, amount of final data and quality of results. This operational flow is sequentially repeated to cover the 17 exposures (4 spectro + 12 photo + 1 dark) to be performed during each single cycle.
At the end of the pipeline, final generated data, with their associated header and metadata to properly re-construct images on ground, are transmitted to the spacecraft Mass Memory Unit, to be down-linked to ground.

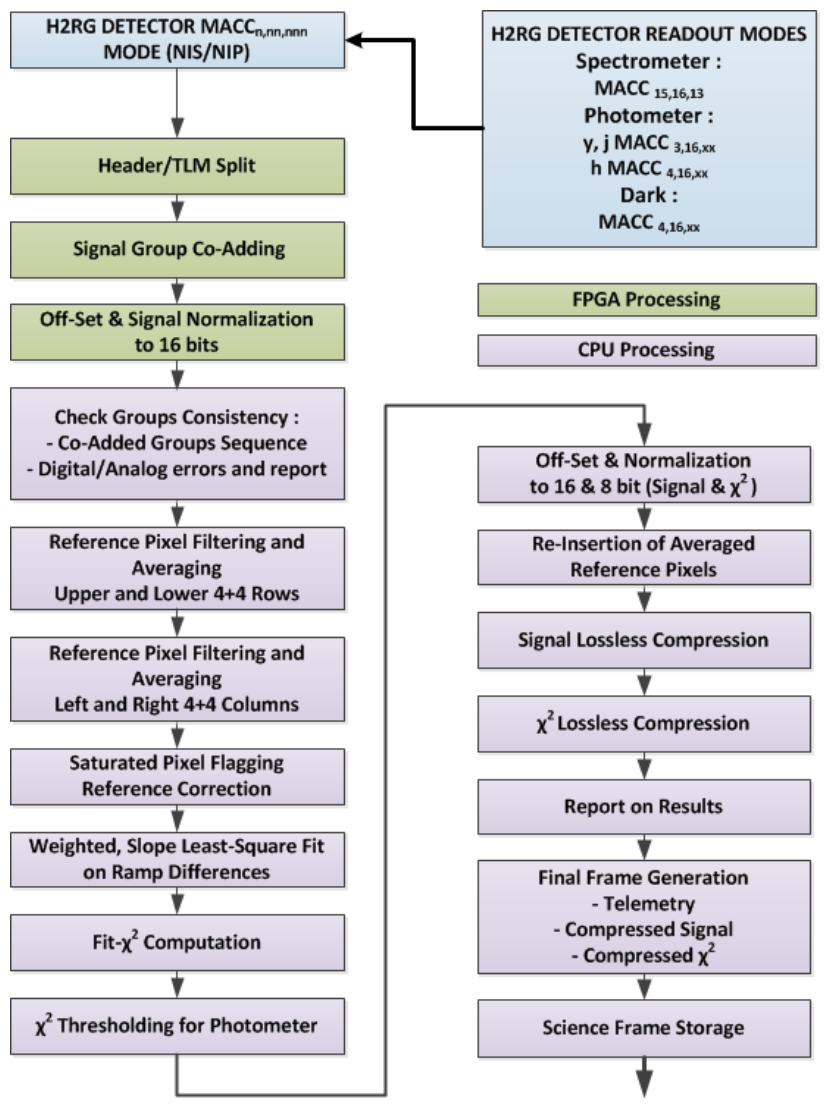

Figure 18: On-board data processing pipe-line for the Euclid NIS/NIP instrumental modes. The pipe-line is subdivided in three different sections on the base of the involved hardware, in the order: SCE analog hardware, FPGA hardware and sequential processing hardware

## 7. NISP MODELS AND DEVELOPMENT

Four NISP models have been developed.

**STM (Structural and Thermal Model):**

The objective of this model was to validate the design of the NI-OMA & NI-DS structure and thermal control by doing the vibration and the TB/TV tests

This model has been successfully tested.

During vibration, the STM model behavior was as expected, particularly, the measured first frequency is very close to the prediction.

During TB/TV, the STM model behavior was as expected, particularly, the gradients in the structure and between the different components are as expected.

This STM model has been delivered to ESA (2017).

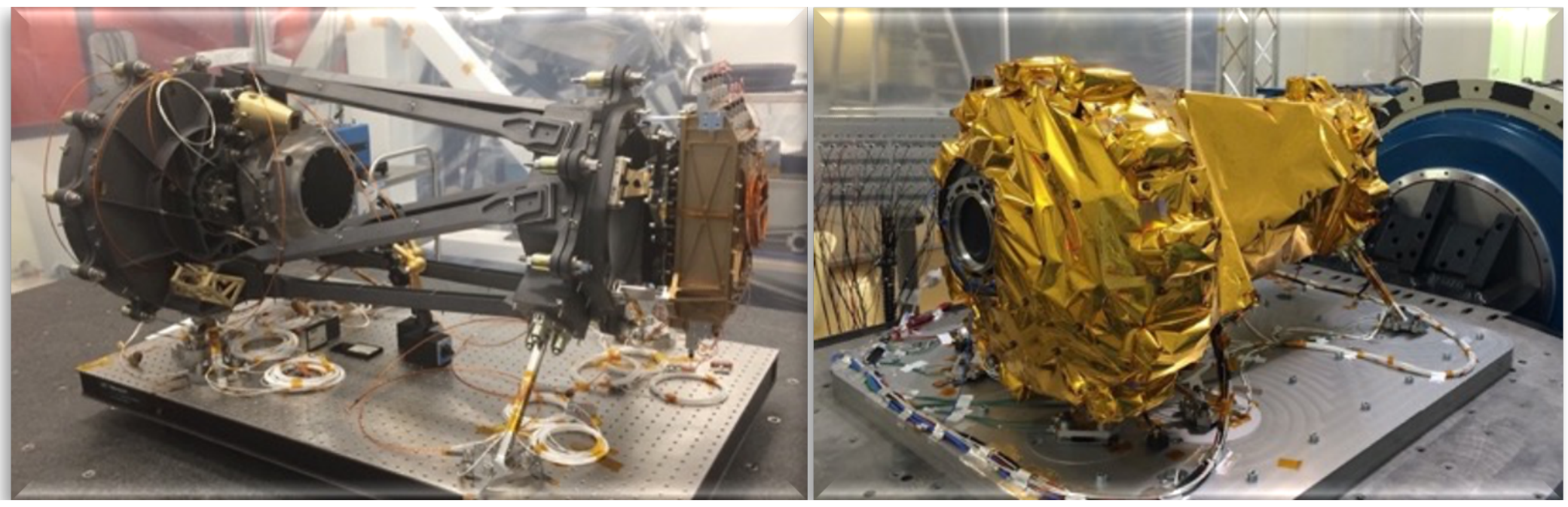

Figure 19: NISP STM model

**Engineering Model (EM):**

An engineering model of NISP is built with all the NISP subsystems qualification model (QM) excepted the structure and the optics. Flight representative harness interconnected the NI-ICU, NI-DPU, NI-DS, NI-TC, NI-FWA, NI-GWA and NI-CU as in flight. The NISP EM has been tested under vacuum at cold operational temperature (135 K for CU, GWA and FWA; [85 K–100 K] for the detector and 140 K for the sidecar electronics.

The purpose of this model is to qualify the functional behavior of NISP (only the nominal side; no redundancy) at cold operational temperature, to perform EMC conducted susceptibility and emission and to prepare the full NISP TV performance to be done on the NISP FM.

During this test we did not succeed to determine the FWA and GWA rotation position using the home search procedure while this procedure was working well at ambient temperature.

This model is not delivered to ESA and will remain in a NISP institute (INAF) as a model for flight software maintenance during flight.

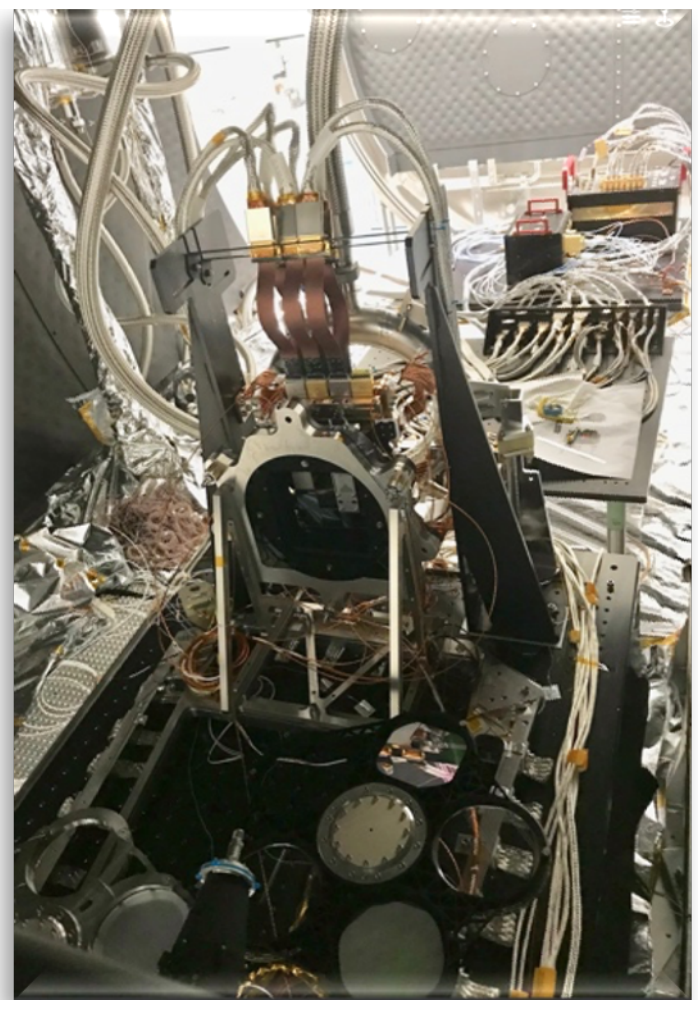

Figure 20: NISP EM model (TV test)

**An Avionic model (AVM):**
A NISP Avionic Model is built with a DPU Engineering model, an ICU Engineering model, a NI-OMA electrical simulator and a SCE Engineering model. It has no redundancy. The objective of the AVM is to test, at warm temperature, the NISP functional performance, the commandability (similar to flight sequence of command), the science data and housekeeping data production.
The NISP AVM has been delivered to ESA and used for the satellite-level test campaign. This model will be maintained by ESA Mission Operation Center during flight.

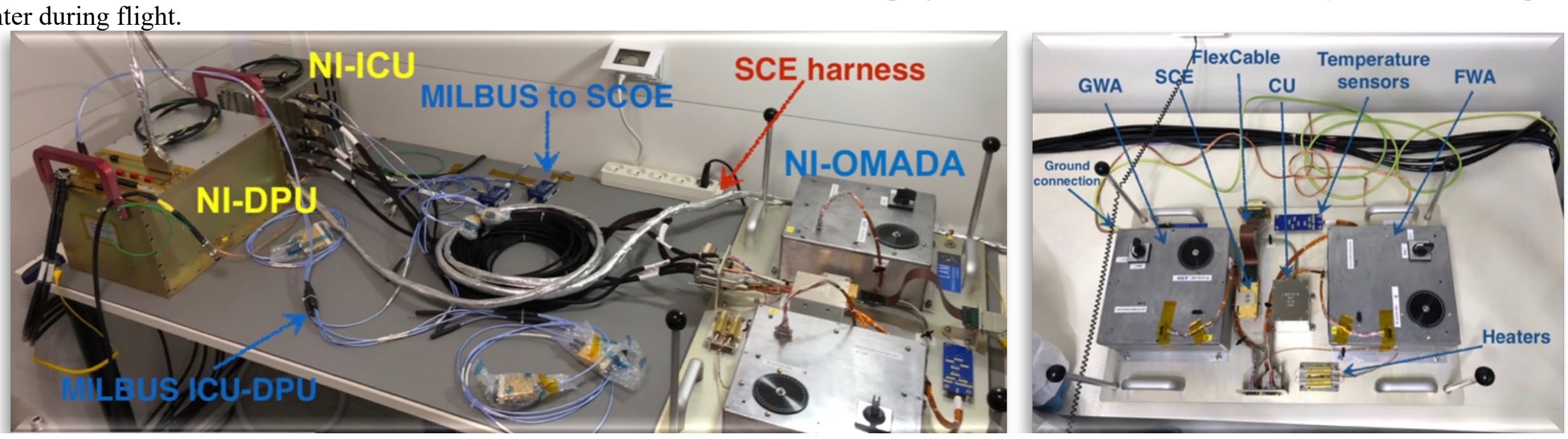

Figure 21: NISP AVM model

**The NISP FM model (FM):**
See description in Sections 3 to 6.
It has been delivered to ESA in May 2020. Then it has been integrated on the Euclid Payload Module (Airbus).

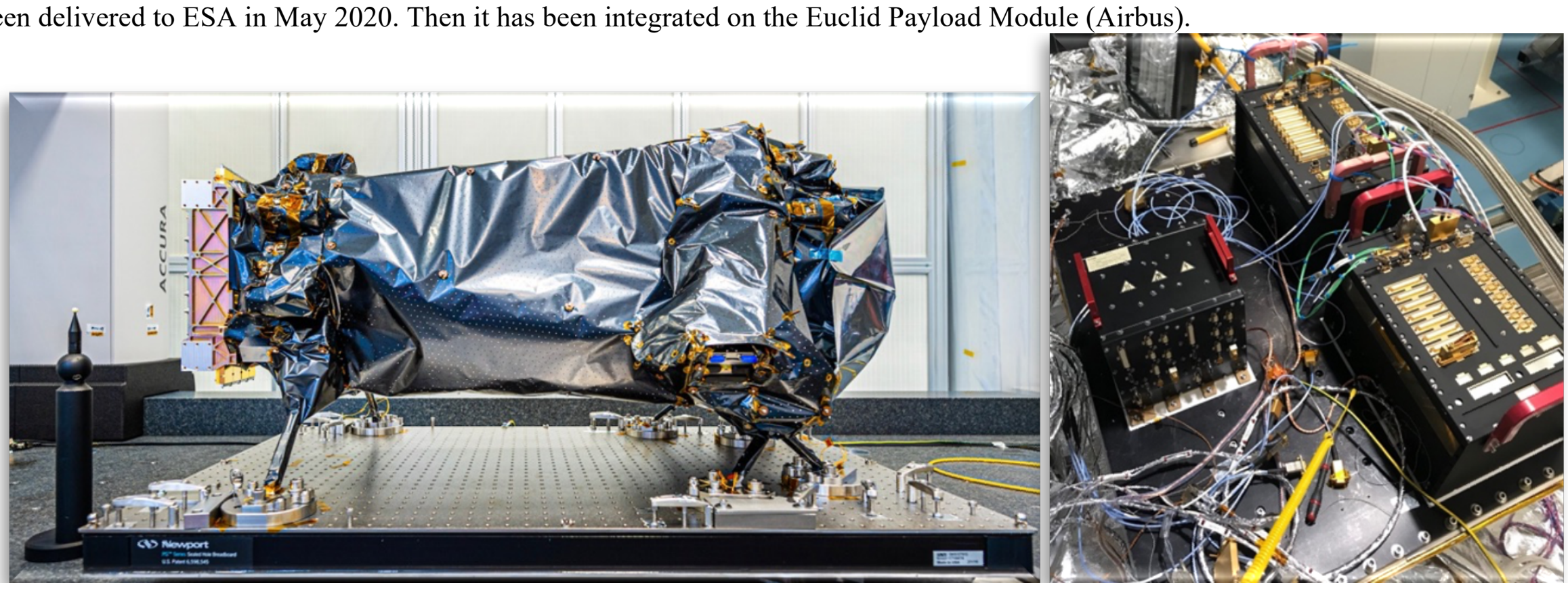

Figure 22: NISP FM model

## 8. NISP TEST AND PERFORMANCE SUMMARY WITH THE FLIGHT MODEL

**Vibration:**

First of all, a successful vibration test has been done at Centre Spatial Liège (CSL).

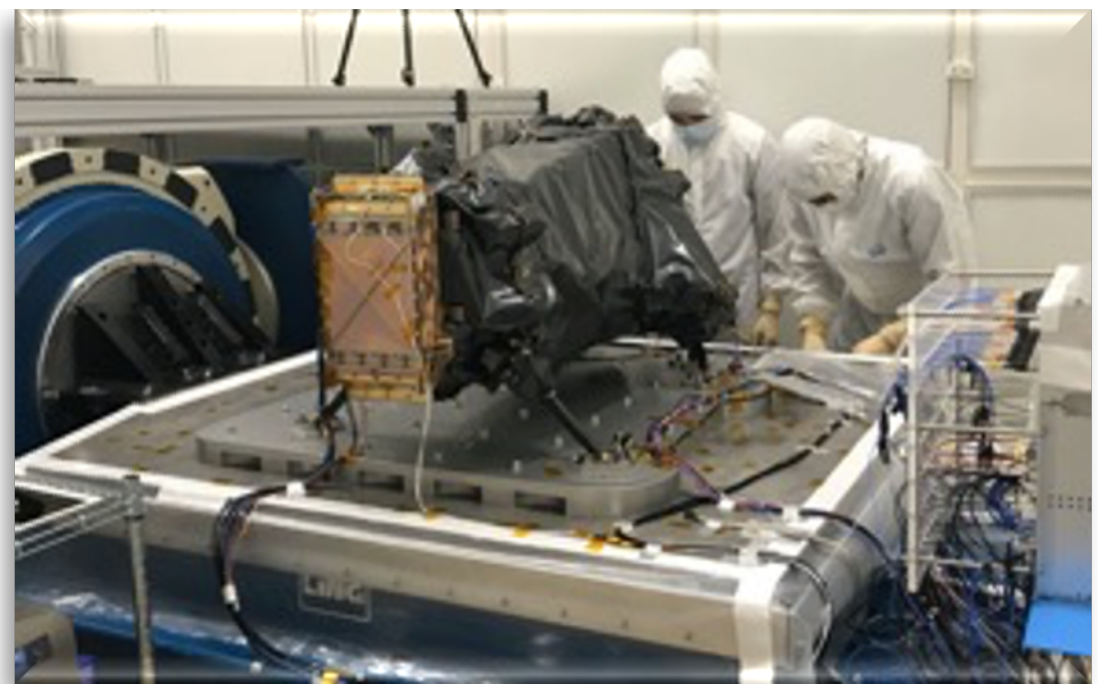

Figure 23: NISP Opto Mechanical Assembly FM during vibration

**First cold thermal vacuum test:**

Then, a first cold thermal vacuum test, has been done, at Laboratoire d'Astrophysisque Marseille (LAM) with the Opto Mechanical Assembly and Detection System, the ICU, and an EGSE (called Markury electronic) replacing the DPUs.

This test was originally not expected. It has been decided, for schedule delay minimization, following the decision to replace the model of the detector sidecar electronic by new electronic design. The characterization of each individual detector triplet (detector chip, flex and sidecar electronic) already done with the first issue of the sidecar "flight" electronic was no more fully valid with the new version of the real flight electronic. In order to minimize the schedule delay, the characterization of the 16 detectors/electronic in a single test has been decided.

The objective of this test were:

- Personality file for each detector, ready for Flight
- NI-CU Illumination LUT
- SCS Noise Characterization
- SCS Non-linearity Correction LUT with Illumination
- SCS Electrical Transfer Function
- Illumination Reference Data with Slew in Survey

This was a very intensive test of 30 days (24h/24h).

More than 100 TBytes of data have been acquired. All data to get very good detector characterization were obtained.

Despite many investigations since the EM TV test, we failed anymore to determine the FWA and GWA position using the home search procedure while this procedure was working well at ambient temperature for the NISP FM. However, we have developed procedures, based on image analysis, to determine the correct position of the two wheels.

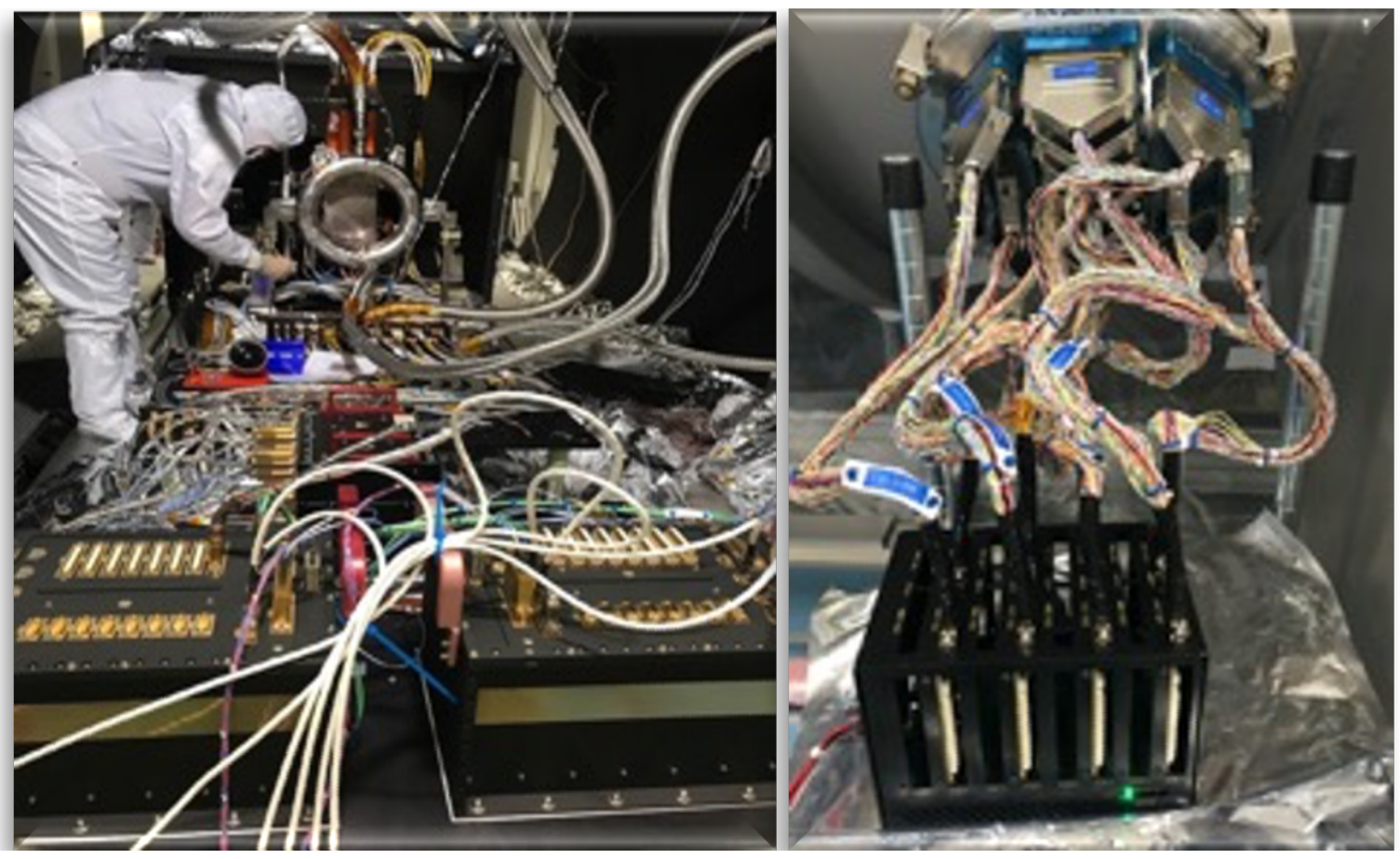

Figure 24: NISP Opto Mechanical Assembly FM during TV N°1 and Markury Electronic (EGSE)

**Second and final cold thermal vacuum test for performance and calibration:**
For this test, done at LAM, the NISP FM was in its complete flight configuration. In order to measure the optical performance, a Euclid telescope simulator was placed in front of the NISP [13]. It was possible to illuminate the NISP optical field with different sources (monochromators, Fabry-Perot etalon, and white lamp). The harnesses between the warm electronic and the cold NISP part was a “flight equivalent” harness provided by ESA.
The main objective of this test is:

- to validate the full NISP functionalities in its operational temperature range
- to determine the NISP focus plane characteristics
- to measure the NISP optical quality (encircled energy of the PSF for all NISP observation modes)
- to get NISP spectroscopic calibration
- to determine the NISP Signal-to-Noise-Ratio performance
- to verify that there is no in- & out-of-field straylight issue

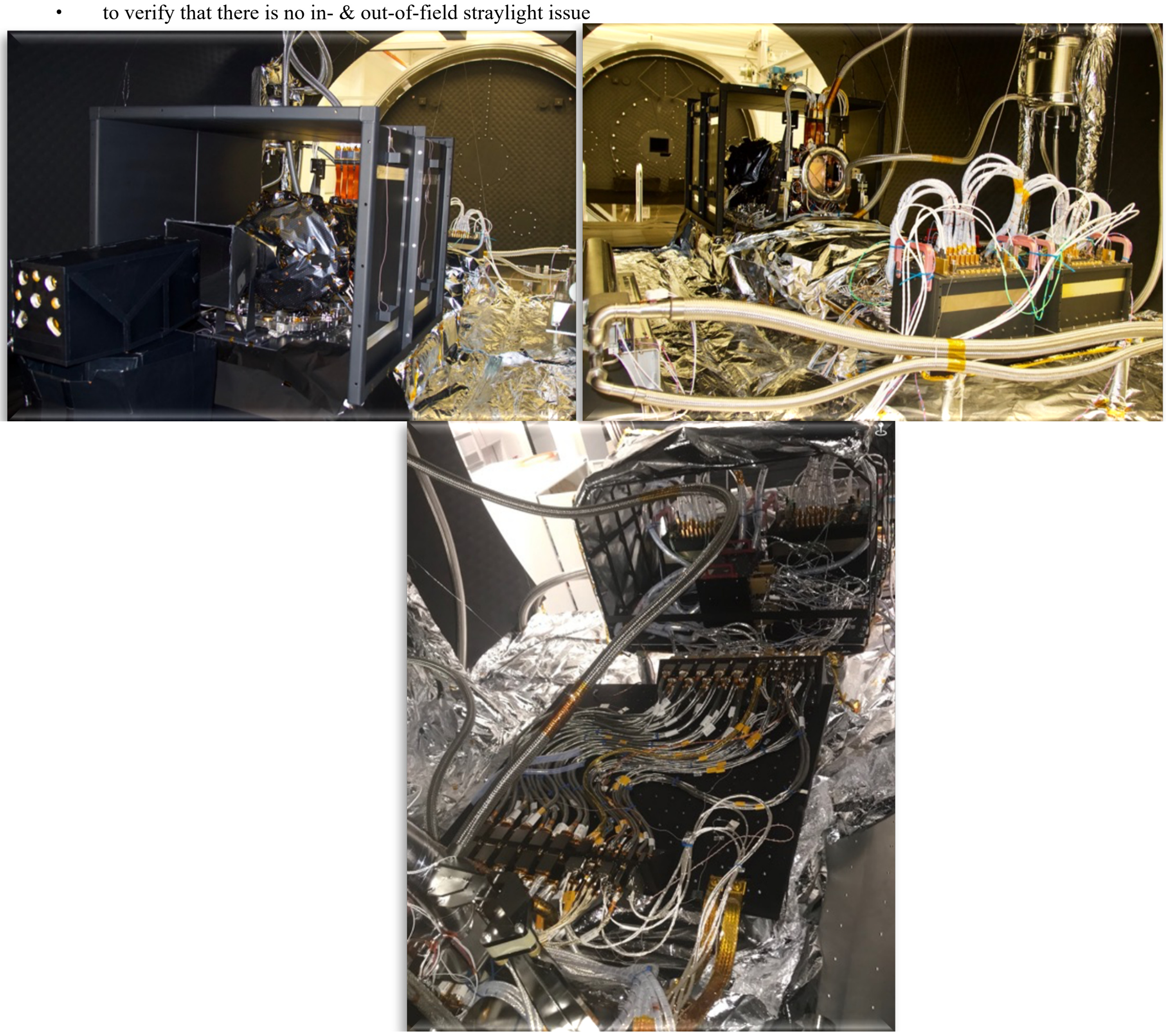

Figure 25: NISP FM during the final TV performance test

This was a very intensive test during 32 days (24h/24h). More than 300000 detector images have been acquired.
The main outcome of this test is that the NISP FM performance has been demonstrated to be better than the specifications (encircled energy, SNR, spectral performance) and that the focus is as expected except for RGS270.

***Encircled energy***
An example of encircled energy measurement shows that the image quality is close to a perfect optical system.

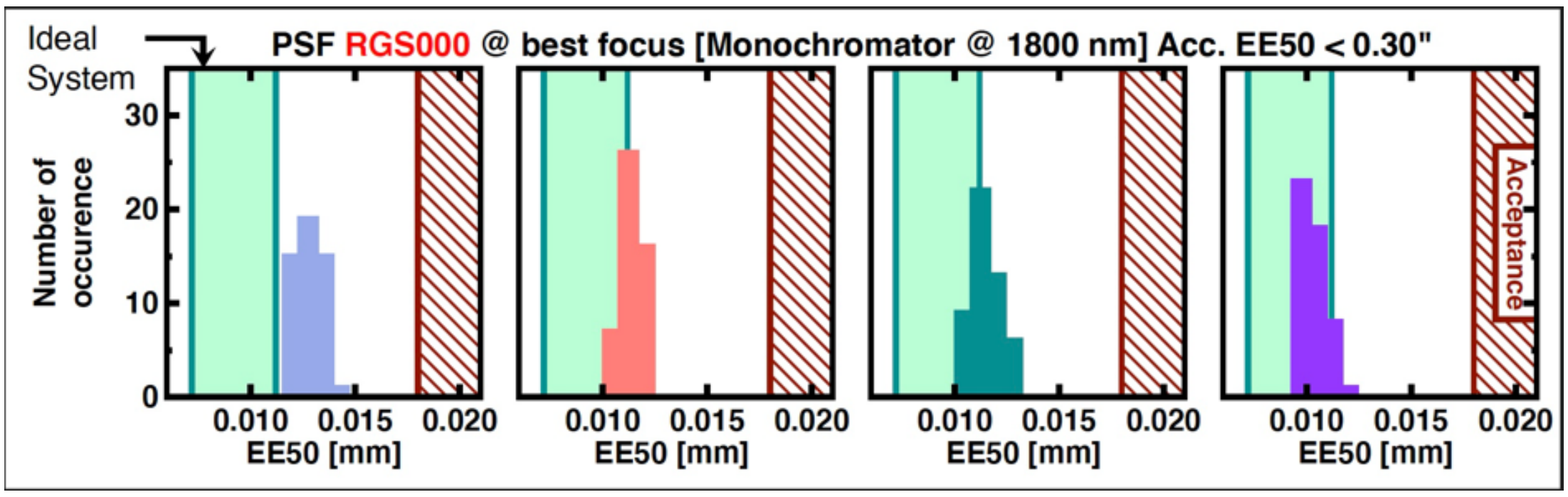

Figure 26: an example of PSF encircled energy

***Focus:***
The focus measured over five different detectors is as expected.

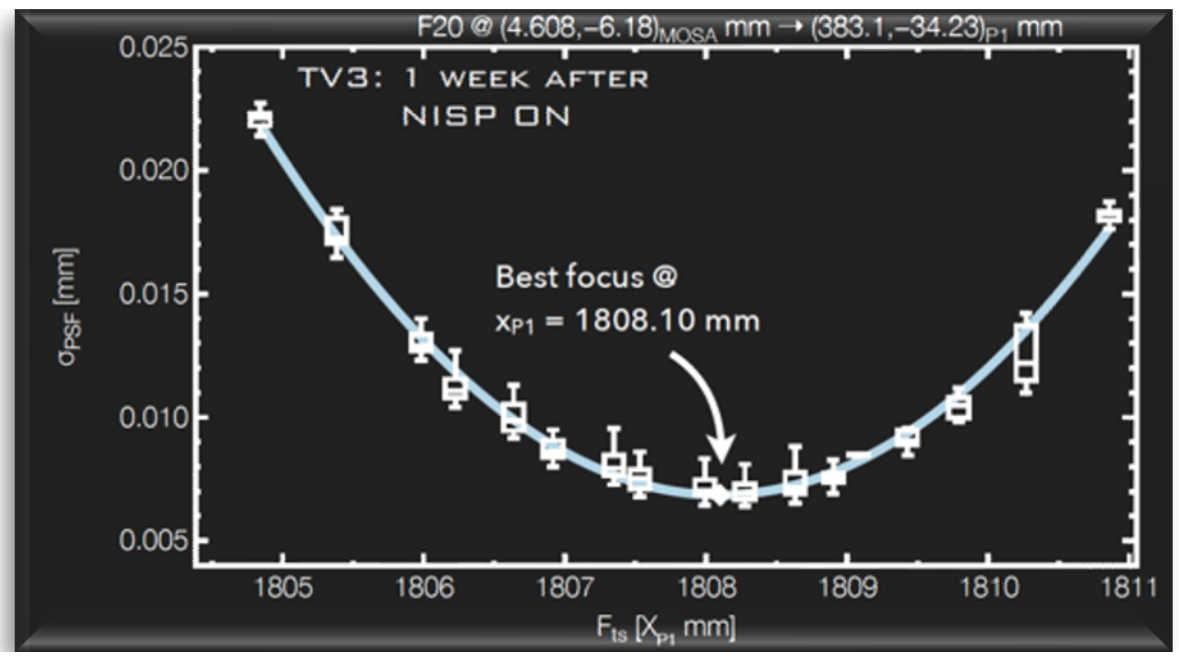

Figure 27: an example of focus measurement

**SNR:**
The estimated Signal-to-Noise is better than specification (5 in photo & 3.5 in spectro for reference sources)

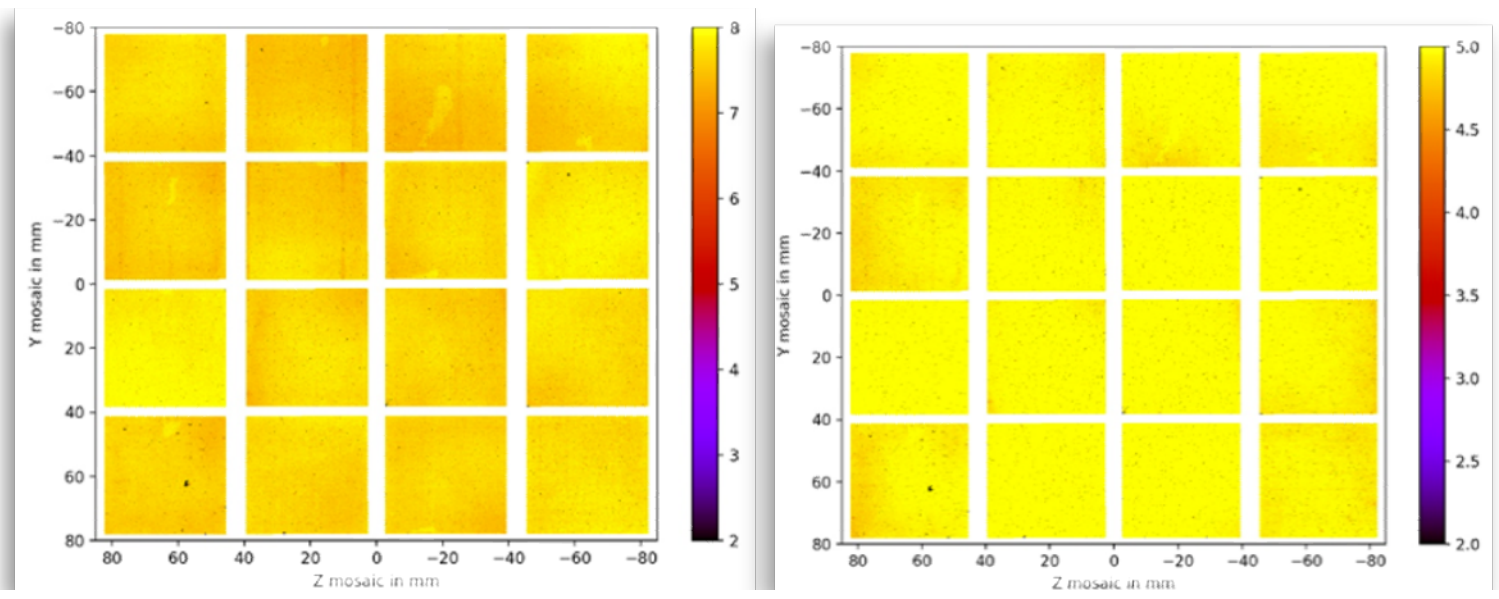

Figure 28: an example of SNR estimation

***SPECTRAL CALIBRATION:***
High optical quality provides spectra of high quality for the spectral calibration

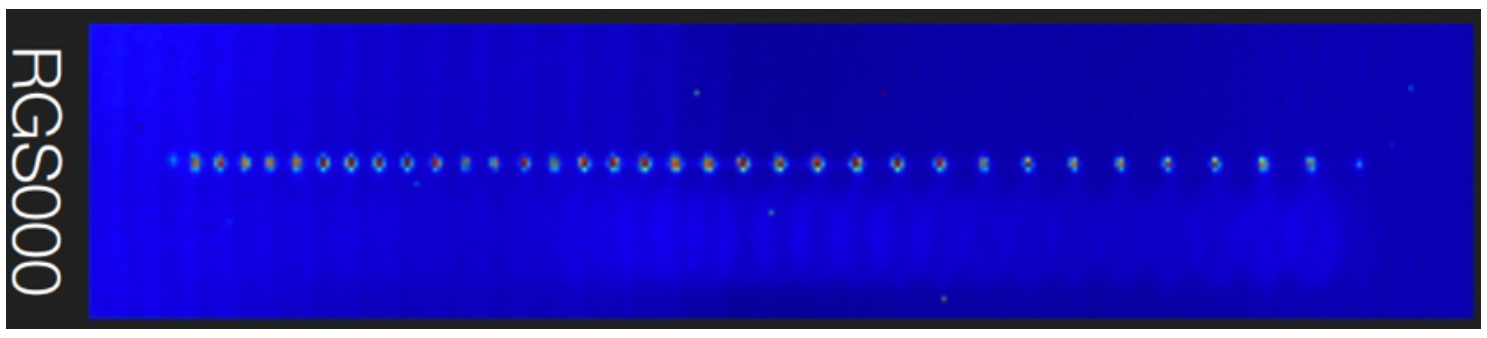

Figure 29: an example of spectrum with etalon

A major issue was found for Grism 270°; the PSF is sharp and within requirement on the blue side but it is strongly degraded with increasing wavelength

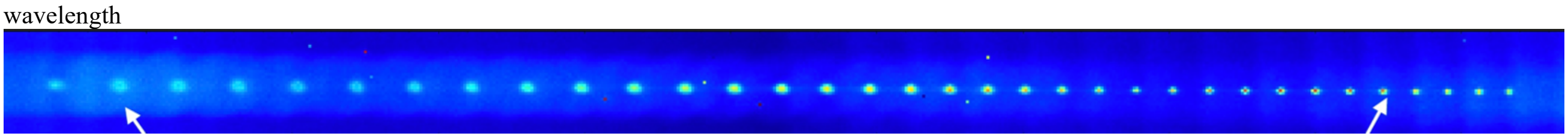

Figure 30: RGS 270 good PSF in the blue part (right) but very bad PSF in the red part (left)

The issue is due to a mismatch between mechanical and optical reference frame in the manufacturing design, which resulted in a wrong realization. After the test, a tiger team has been put in place. The conclusion is that the Grism 270° will be discarded from the observation, but the Grism 0° and the Grism 180° will be used with nominal and +/– 4° of rotation (see the NISP sequence observation in §2 and [14]). During NISP ground test, it has been demonstrated that using the Grism 0º and 180º at +/–4° from their nominal position has no impact on the spectral performance. The new sequence of observation, without the Grism 270°, have equivalent completeness and purity performance. Therefore, no Grism 270° remanufacturing and replacement was implemented.

***NISP performance during the Euclid PLM cold TV test:***

After the integration and the alignment of the NISP FM on the Euclid PLM (Airbus), after EMC and vibration tests, a PLM cold performance test has been done.

Unfortunately, only 7 detectors where operational during this test. The issue was understood after the TV test. It is a conjunction of a very particular FM harness + test extension harness length that has triggered a DPU ASW bug never seen before despite many tests done with many different harness lengths. DPU "FIFO error" can be present for a very specific delay on the LVDS line between DPU and detector electronic, and the range of delay is very small (<3ns). Before the DPU bug correction, the DPU ASW mechanism – treating this "error", that should indeed not be considered, as a major error -- was not working properly. This has been demonstrated during a specific TV test done with DPU and detector electronic EQM and different harness lengths. DPU ASW has been modified, tested and loaded on DPU FM's. This issue is completely solved.

Using only the 7 working detectors in the center and the corner of the field, the test results show that NISP is well aligned with VIS and produces high quality images once the Euclid telescope is focused on VIS.

The NISP PSF encircled energy is as expected following the NISP TV tests and better than the scientific requirement

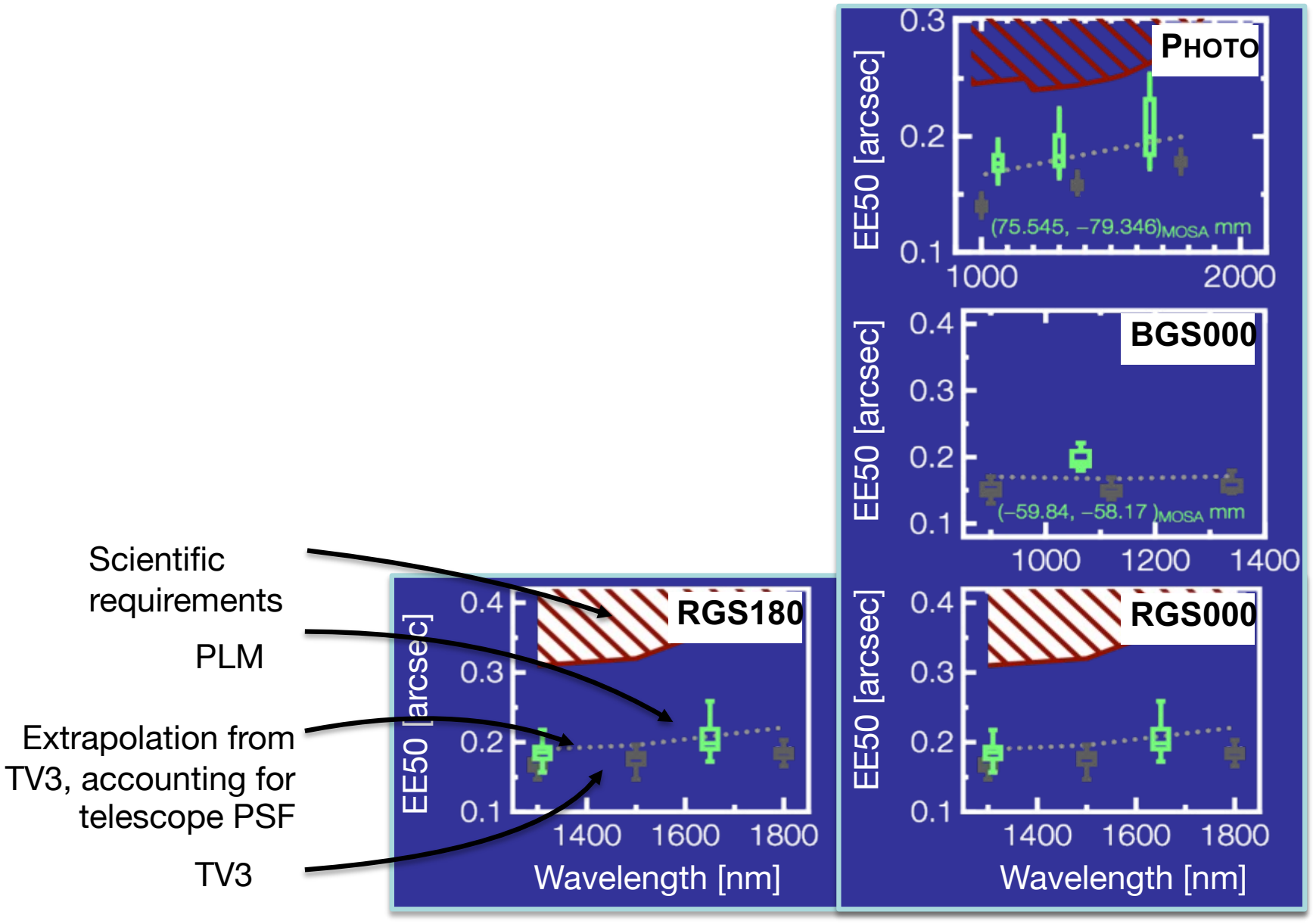

Figure 31: an example of PSF encircled energy obtained during Euclid PLM TV test

During this test, the determination of the FWA and GWA position, at cold, using the home search procedure was working well for NISP nominal side but not for NISP redundant side. We have validated that the wheel position determined with image analysis is also working well at PLM level. Considering that the FWA and GWA position using the home search procedure malfunctioning is not understood (this is a global system ICU/Harness/wheel home sensor), the procedure using image analysis to determine the FWA and GWA correct position is proposed to ESA as nominal procedure for flight.

## 9. CONCLUSION

The NISP FM performance has been demonstrated to be better than the specifications (encircled energy, SNR, spectral performance), at NISP level and at Euclid PLM level except for RGS270. A new sequence of observation, without the Grism 270°, has equivalent completeness and purity performance than the originally planned sequence with the Grism 270°.

The NISP team is (by mid 2022) actively working in preparing the operations and support the System Verification Operational Test managed by ESA's Mission Operations Center (MOC).

The full Euclid satellite is now fully integrated and is about to start its nominal sequence of tests (no optical performance are expected to be done at satellite level).

The Euclid launch readiness is expected for mid 2023.

## ACKNOWLEDGMENTS

The Euclid Consortium acknowledges the European Space Agency and a number of agencies and institutes that have supported the development of Euclid, in particular the Academy of Finland, the Agenzia Spaziale Italiana, the Belgian Science Policy, the Canadian Euclid Consortium, the Centre National d'Etudes Spatiales, the Deutsches Zentrum für Luft und Raumfahrt, the Danish Space Research Institute, the Fundação para a Ciência e a Tecnologia, the Ministerio de Economia y Competitividad, the National Aeronautics and Space Administration, the National Astronomical Observatory of Japan, the Netherlandse Onderzoekschool Voor Astronomie, the Norwegian Space Agency, the Romanian Space Agency, the State Secretariat for Education, Research and Innovation (SERI) at the Swiss Space Office (SSO), and the United Kingdom Space Agency. A complete and detailed list is available on the Euclid web site (http://www.euclid-ec.org).